\documentclass[12pt,letterpaper]{article}
\usepackage{authblk}
\usepackage[intlimits]{amsmath} 
\usepackage{amsxtra}     
\usepackage{amsthm}
\usepackage{amssymb}
\usepackage{float}
\usepackage{amsmath}
\usepackage{bm}% bold math
\usepackage{amsfonts}
\usepackage{graphicx}    
\usepackage{rotating}
\usepackage{color}
\usepackage{cite}
\usepackage{xcolor}
\usepackage{epsfig}
\usepackage{subfigure}
\usepackage{hyperref}       % hyperlinks
\usepackage{url}            % simple URL typesetting
\usepackage{booktabs}       % professional-quality tables
\usepackage{amsfonts}       % blackboard math symbols
\usepackage{setspace}
\usepackage{multirow}
\usepackage{longtable}
\usepackage{supertabular}
\usepackage{sectsty}
\usepackage[title]{appendix}
\usepackage[margin=1.4in]{geometry}

\newcommand{\angstrom}{\mbox{\normalfont\AA}} %for angstrom symbol
\newcommand{\tabincell}[2]{\begin{tabular}{@{}#1@{}}#2\end{tabular}} % Force line break in a cell of a table
\begin{document}

\title{Screening of generalized stacking fault energies, surface energies and intrinsic ductile potency of refractory multicomponent alloys}
\author[1]{Yong-Jie Hu}
\author[1]{Aditya Sundar}
\author[2]{Shigenobu Ogata} 
%\author[3]{William A. Curtin}
\author[1]{\\ Liang Qi\thanks{Corresponding author: qiliang@umich.edu}}

\affil[1]{Department of Materials Science and Engineering,
  University of Michigan, Ann Arbor 48109, USA}
\affil[2]{Department of Mechanical Science and Bioengineering, 
Osaka University, Osaka 560-8531, Japan}
%\affil[3]{EPFL}
\date{}
\hyphenpenalty=5000
%\righthyphenmin=62
\maketitle

\pagebreak

\begin{abstract}
 Body-centered cubic (bcc) refractory multicomponent alloys are of great interest due to their remarkable strength at high temperatures. Meanwhile, further optimizing the chemical compositions of these alloys to achieve a combination of high strength and room-temperature ductility remains challenging, which would require systematic predictions of the correlated alloy properties across a vast compositional space. In the present work, we performed first-principles calculations with the special quasi-random structure (SQS) method to predict the unstable stacking fault energy ($\gamma_{usf}$) of the $(1\bar10)[111]$ slip system and the $(1\bar10)$-plane surface energy ($\gamma_{surf}$) for 106 individual binary, ternary and quaternary bcc solid-solution alloys with constituent elements among Ti, Zr, Hf, V, Nb, Ta, Mo, W, Re and Ru. Moreover, with the first-principles data and a set of physics-informed descriptors, we developed surrogate models based on statistical regression to accurately and efficiently predict $\gamma_{usf}$ and $\gamma_{surf}$ for refractory multicomponent alloys in the 10-element compositional space. Building upon binary and ternary data, the surrogate models show outstanding predictive ability in the high-order multicomponent systems. The ratio between $\gamma_{surf}$ and $\gamma_{usf}$ is a parameter to reflect the potency of intrinsic ductility of an alloy based on the Rice model of crack-tip deformation. Therefore, using the surrogate models, we performed a systematic screening of $\gamma_{usf}$, $\gamma_{surf}$ and their ratio over 112,378 alloy compositions to search for alloy candidates that may have enhanced strength-ductile synergies. Search results were also confirmed by additional first-principles calculations.
\end{abstract}

\pagebreak

\section{Introduction}
\label{section:intro}
Metals and alloys based on transition metal elements of Group V and VI in the periodic table (such as Nb, Ta, Mo, and W) usually have nearly half-filled d-band electrons. These electrons generate strong interatomic bonds with considerable directional dependence, and they make these alloys have the non-close packed bcc lattice structure\cite{sutton1993electronic}. Because of these electronic and atomistic structures, these alloys have high melting temperatures, and the dislocation motions in these alloys have more substantial activation barriers and temperature-dependent behavior compared with many other metallic alloys\cite{brunner2000comparison,takeuchi1977slip,suzuki1999plastic,dezerald2014ab,savitskii2012physical}. Thus, these alloys (so-called refractory metals and alloys) have excellent mechanical performances with high strengths and sufficient ductility at high temperatures ($>$1000 $^\circ$C), but many of them are brittle at the room temperature, significantly limiting their mechanical processing and engineering applications.

Recently, there are developments of multicomponent alloys based on Group IV, V and VI elements mixed in either equimolar and non-equimolar ratios\cite{senkov2010refractory,senkov2018development,soni2020phase,lee2018lattice,gao2016design,gao2017computational}. Some of these refractory high-entropy alloys (HEAs) were reported to have excellent mechanical performances in both strengths and ductility in extremely high-temperature regions\cite{senkov2010refractory,senkov2019high,senkov2018development,yao2017mechanical,sheikh2016alloy,coury2019solid,shi2019enhanced}. However, these alloys generally have low ductility at room temperatures\cite{senkov2018development,zhang2014microstructures}. Because there are many degrees of freedom in the compositional space, it is urgent to develop efficient and accurate methods to predict the strengths and ductility of candidate alloys with arbitrary chemical compositions in order to search for alloys with optimized mechanical performances. So far, there have been many studies to predict the strengths of multicomponent HEA in both face-centered cubic (fcc) and bcc lattice structures\cite{rao2019solution,varvenne2016theory,yin2020vanadium,maresca2020mechanistic,george2020high,larosa2019solid,rao2017atomistic}. However, only sparse theoretical studies were conducted to investigate their ductility, especially for bcc HEAs\cite{sheikh2016alloy,li2020ductile}.

The evaluations of the ductility for general solid solution alloys using first-principles calculations are often conducted based on several well-established criteria. They include the Pugh’s modulus ratio of the bulk and shear modulus of alloys\cite{pugh1954xcii}, the lattice instability mechanism under ideal strength deformation\cite{qi2014tuning,yang2018ab,winter2019intrinsic}, and the Rice criterion based on the competition between dislocation emissions and cleavage fracture propagation\cite{rice1974ductile,rice1992dislocation}. Using the linear elastic fracture mechanics (LEFM) analyses, the critical stress intensity factor for cleavage fracture propagation near a crack tip can be predicted by using elastic constants ($c_{ij}$) and surface energies ($\gamma_{surf}$) of cleavage planes, and the critical stress intensity factor for dislocation emissions near the crack tip can be predicted by using elastic constants and unstable stacking fault (USF) energies ($\gamma_{usf}$) of specific slip systems. The intrinsic ductility of an alloy can be determined based on the ratio between these two types of stress intensity factors. Approximately, an alloy with a higher ratio of $\frac{\gamma_{surf}}{\gamma_{usf}}$ can be considered to have a stronger potency of being intrinsic ductile\cite{rice1974ductile,rice1992dislocation,li2020ductile,wu2015brittle}. These approaches have been applied using either empirical interatomic potentials or first-principles calculations for different metals and alloys\cite{andric2017new,zhu2004atomistic,wu2015brittle,li2020ductile}, including bcc HEAs\cite{li2020ductile}.

However, to calculate these parameters ($c_{ij}$, $\gamma_{surf}$ and $\gamma_{usf}$) of multicomponent solid-solution alloys based on first-principles calculations are not straightforward tasks. These calculations were often conducted using the relatively large supercells generated by the special quasi-random structure (SQS) method\cite{VanDeWalle2013}, which tunes the correlation functions of lattice occupations in the finite-size supercells to be close to those of the ideally mixed solid solutions. Multiple first-principles density functional theory (DFT) calculations have to be conducted to obtain the average results of the parameters for a specific alloy composition\cite{de2016calculations,DeJong2015}. These expensive calculations limit our ability to explore compositional spaces efficiently.

Statistical learning methods can be applied to construct surrogate models to predict the parameters of HEAs and other multicomponent solid-solution alloys with DFT-level accuracies. However, a key bottleneck of these surrogate models is the small size of training data sets intrinsically limited by the costs of DFT calculations, which can undermine their extrapolative prediction ability. This limitation could be relieved by including the physical mechanisms in the surrogate model. As discussed above, the atomistic and electronic structures are strongly correlated to the deformation defect properties and the corresponding mechanical behavior of bcc transition-metal alloys. For example, the bcc alloys based on Group V elements are generally ductile, but those based on Group VI elements are generally brittle, although the latter have higher strengths, and these variations are controlled by the average filling level of d-band electrons\cite{qi2014tuning,yang2018ab}. Our recent studies also reveal the stability of deformation defects in bcc transition-metal alloys are strongly correlated to the features of local d-band shape and filling level\cite{hu2019local}. Thus, it is possible to combine statistical regression methods and feature parameters of electronic and atomistic structures to construct an accurate and reliable model for bcc solid-solution alloys.

In this paper, we developed surrogate models based on statistical regression to learn the DFT calculations of USF energies ($\gamma_{usf}$) and surface energies ($\gamma_{surf}$) of $\{$110$\}$ plane in multicomponent bcc solid-solution alloys mainly composed of Group IV, V, and VI elements. Based on a set of descriptors for capturing the features of atomic bonds and electronic structures of pure metals and ordered intermetallic alloys, our models can successfully predict the variations of USF and surface energies in a large multicomponent space using only the chemical compositions as the inputs. Our current regression model trained only based on $\sim$70 data of binary and ternary alloys can accurately predict $\gamma_{usf}$ and $\gamma_{surf}$ for quaternary alloys different from those in the training data. Using our surrogate models, we conducted a fast screening of $\gamma_{usf}$ and $\gamma_{surf}$ over a large number of quaternary bcc alloys, which are kinetically possible to be synthesized based on the currently available phase diagrams. The predictions of many extreme cases from these screening results were then confirmed by additional DFT calculations. The results suggest that their potency of strengths and ductility, which are related to $\gamma_{usf}$ and $\frac{\gamma_{surf}}{\gamma_{usf}}$, respectively, are not solely determined by the $d$-band filling. In addition, there could be considerable spaces to tune alloy chemical compositions for further improvements of the strength-ductility synergy relative to the currently known equimolar HEAs.

We have to emphasize that several major approximations and simplifications have been taken in our framework. First, we only calculate the USF and surface energies without considering the composition effects on the elastic constants of alloys, which are needed to evaluate the critical stress intensity factor. Second, there could be multiple slip planes and fracture surface planes in bcc alloy systems besides $\{$110$\}$ planes. In principles, we can also predict these parameters based on the same framework of DFT calculations and statistical regression methods as discussed in this paper, but it would heavily increase our calculation efforts. Additionally, the elastic constants of solid-solution alloys can also be derived from recently developed machine learning models\cite{kim2019first}. Other factors, such as the variations of local GSF\cite{ding2018tunable,li2019strengthening,Natarajan2020}, lattice trapping effects\cite{Curtin90crack,Gumbsch03tungsten} and the effects of short-range ordering\cite{li2019strengthening,zhang2020short,singh2018design,Natarajan2020} and finite temperature\cite{Natarajan2020}, cannot be described by the average results of USF and surface energies from the first-principles SQS method. Additionally, the contributions of deformation twining could be important to determine the strength and ductility of some specific HEAs\cite{senkov2012microstructure,couzinie2015room,lilensten2017design}, which can be further investigated in the future by applying the similar DFT methodologies\cite{niu2018magnetically,ogata2005energy,de2016calculations} in the future. Our strategy here is to develop these statistic regression models to efficiently search in the multicomponent compositional spaces to find the possible alloy compositions with optimized values of $\gamma_{usf}$ and $\frac{\gamma_{surf}}{\gamma_{usf}}$. Once such candidates are identified, more rigorous DFT calculations and defect models can be applied to evaluate their mechanical performances.

\section{Methods}
\label{section:methods}
\subsection{DFT calculations}
\subsubsection{Computation of unstable stacking fault and surface energy}
\label{section:USF_surf_method}

In the present work, the supercells used for DFT calculations were built based on the special quasi-random structure (SQS) method\cite{Zunger1990} to approximately describe the chemical disorder in the studied solid-solution alloys. The SQSs were generated by using the Alloy Theoretic Automated Toolkit (ATAT)\cite{VanDeWalle2013}, in which a Monte Carlo-based evolutionary algorithm is used to search the periodic atomic structure with the closest match of correlation functions of a ideally mixed solid-solution state. Here, 13 types of SQSs were generated to study the bcc solid-solution phases with different binary, ternary and quaternary alloying compositions in a 10-element compositional space (i.e., Ti, Zr, Hf, V, Nb, Ta, Mo, W, Re and Ru). As listed in Table \ref{SQS:list}, 64 individual compositions were modeled for 14 binary systems, 10 compositions for 3 ternary systems, and 32 compositions for 12 quaternary systems.  

\begin{table}[!htbp]
    \centering
    \caption{Alloying systems and compositions studied by the SQS method}
    \label{SQS:list}
    \begin{tabular}{cc}
    \\
    %\hline
    \hline
    Systems & Alloying compositions studied by the SQS method \\ \hline
    Binary & \tabincell{c}{
    Ti$_3$Nb, Ti$_2$Nb, TiNb, TiNb$_2$, TiNb$_3$, \\
    Ti$_3$W, Ti$_2$W, TiW, TiW$_2$, TiW$_3$, TiW$_7$, \\
    Ti$_3$Ru, Ti$_2$Ru, TiRu, \\
    Hf$_3$Nb, Hf$_2$Nb, HfNb, HfNb$_2$, HfNb$_3$, \\
    V$_7$W, V$_3$W, VW, VW$_3$, VW$_7$,\\
    Nb$_7$Ta, Nb$_3$Ta, NbTa, NbTa$_3$, NbTa$_7$,\\
    Nb$_3$Mo, Nb$_2$Mo, NbMo, NbMo$_2$, NbMo$_3$,\\
    Nb$_3$W, Nb$_2$W, NbW, NbW$_2$, NbW$_3$,\\
    Nb$_7$Ru, Nb$_3$Ru, Nb$_2$Ru, NbRu, \\
    Ta$_7$Mo, Ta$_3$Mo, TaMo, TaMo$_3$, TaMo$_7$,\\
    Ta$_7$Re, Ta$_3$Re, TaRe, TaRe$_3$, \\
    Mo$_7$W, Mo$_3$W, MoW, MoW$_3$, MoW$_7$,\\
    W$_7$Re, W$_3$Re, WRe, WRe$_3$, \\
    W$_7$Ru, W$_3$Ru, W$_2$Ru
    }      \\ \hline
    Ternary &  \tabincell{c}{
    Ti$_2$NbW, Ti$_2$Nb$_2$W, TiNbW, TiNb$_2$W, TiNb$_2$W$_2$,
    TiNbW$_2$, \\
    TiWRe,TiW$_2$Re, \\
    TiNbRu, TiNb$_2$Ru}     \\ \hline
    Quaternary & \tabincell{c}{
    TiZrHfNb, TiZrHf$_2$Nb$_2$, TiZrHfNb$_3$, \\
    TiZrVNb, TiZr$_2$V$_2$Nb, \\
    TiVNbMo, \\
    TiNbWRe, TiNb$_3$WRe, TiNb$_2$W$_2$Re, TiNbW$_3$Re, \\
    TiNbWRu, TiNb$_3$WRu, TiNb$_2$W$_2$Ru, TiNbW$_3$Ru, \\
    VNbWRu, VNb$_2$W$_2$Ru, VNbW$_3$Ru, \\
    VMoWRu, VMoW$_3$Ru, \\
    Nb$_2$Ta$_2$MoW, NbTaMoW, NbTa$_3$MoW, NbTaMoW$_3$ \\
    Nb$_3$TaWRu, Nb$_2$TaW$_2$Ru, \\
    NbMoW$_3$Ru, \\
    Ta$_2$MoW$_2$Re, TaMoWRe, TaMoW$_3$Re, \\
    TaMoWRu, TaMo$_2$W$_2$Ru, TaMoW$_3$Ru
    }   \\ \hline
    \end{tabular}
    
\end{table}

\begin{figure}[!ht]
    \centering
    \includegraphics[width=1.0\linewidth]{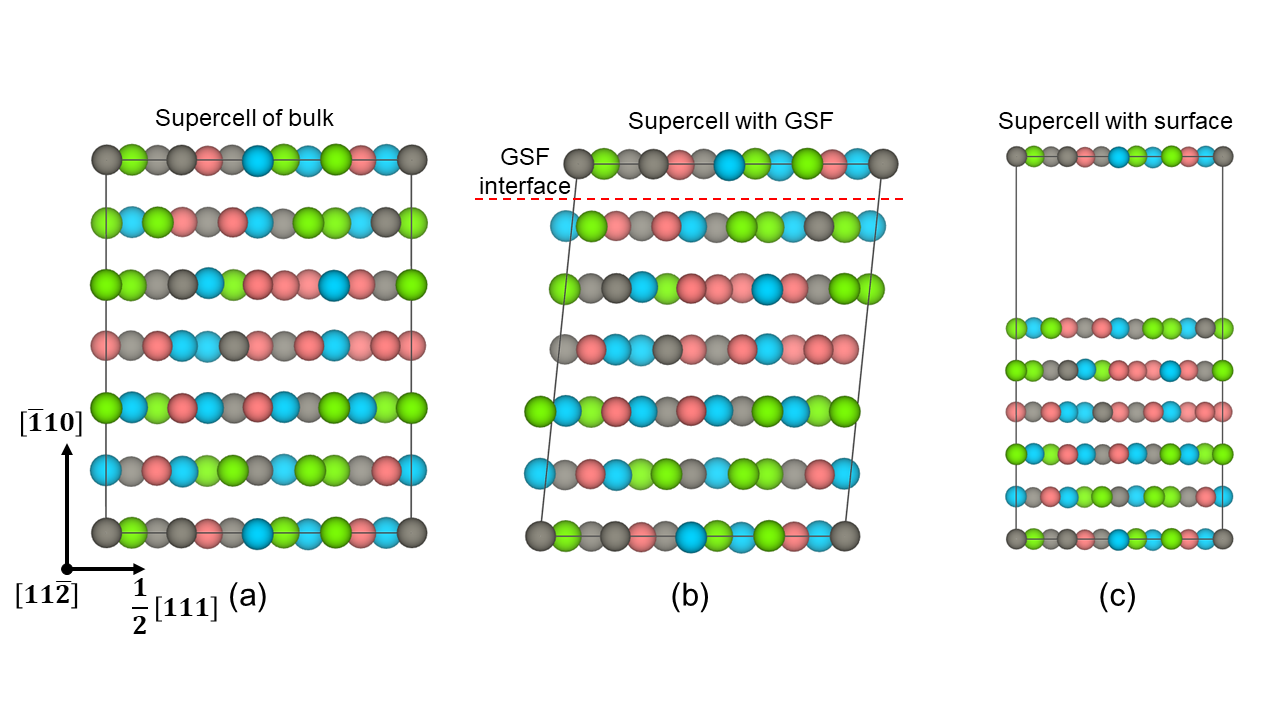}
    \caption{The configurations of (a) the bulk SQS supercell for the quaternary equimolar composition, (b) the supercell for calculations of USF energies, and (c) the supercell for calculations of surface energies. These figures show a projection along the $[11\bar2]$ direction.}
    \label{fig:SQS:cell}
\end{figure}

Most of the generated SQSs are 72-atoms supercells are with orthogonal basis vectors, which are $[11\bar2]\times2[111]\times3[1\bar10]$ presented using the conventional bcc lattice basis. An example of the 72-atoms SQSs is shown in Fig. \ref{fig:SQS:cell}a for a quaternary alloy with equimolar composition. Besides, a SQS with 90 atoms and basis vectors of $[11\bar2]\times2[111]\times5[1\bar10]$ was generated to model the ternary alloys with the A$_2$B$_2$C type of compositions (A, B, and C represent the alloying elements). Moreover, other than the SQSs associated with the compositions listed in Table \ref{SQS:list}, we also generated a additional group of 72-atoms quaternary SQSs in order to validate the screening results of the surrogate models.    

The atomic positions and geometry of the generated SQS supercells are firstly fully relaxed to capture the local lattice distortion induced by the size mismatch between the constituent elements. Then, with the relaxed supercells, a method developed in our previous work\cite{de2016calculations} is adopted to compute the $(1\bar10)[111]$ unstable stacking fault energy, $\gamma_{usf}$, and the $(1\bar10)$ surface energy, $\gamma_{surf}$.

To create a $(1\bar10)[111]$ generalized stacking fault (GSF) between two neighboring $(1\bar10)$ planes, the atoms below the fault interface are rigidly shifted relative to the rest of atoms along the $[111]$ direction. The shift can be accommodated by distorting the bulk supercell (Fig. \ref{fig:SQS:cell}a) to have an angle away from 90 degrees between the $[111]$ and $[\bar110]$ directions so that only one stacking fault interface is introduced in the supercell, as shown in Fig. \ref{fig:SQS:cell}b\cite{DeJong2016}. The shifted supercell is then relaxed to compute the total energy under certain constraints: (1) all the atoms are relaxed along the $[\bar110]$ direction but fixed along the rest of two directions; (2) the supercell size along $[\bar110]$ direction is also relaxed to remove the normal stress perpendicular to the fault plane. The GSF energy is calculated based on the total energy increase caused by the shift relative to the undeformed structure. 

In the present work, for sake of simplification, the USF energy ($\gamma_{usf}$) is treated as the GSF energy at a fixed shift distance that equals to the length of $\frac{1}{4}[111]$, although a more rigorous way is to interpolate the maximum point of the GSF energy curve. Our benchmark calculations were performed to show that the difference between the GSF energy at $\frac{1}{4}[111]$ and the maximum energy from curve interpolation is actually negligible. Details are shown in Section \ref{section:verifi_comp_appro}. In addition, since one supercell contains multiple $(1\bar10)$ planes, the GSFs between all the possible neighboring planes must be considered. For each of the two neighboring planes, two shifts, namely $-\frac{1}{4}[111]$ and $+\frac{1}{4}[111]$, have been applied to compute the corresponding GSF energies, respectively. Finally, for a given alloy composition, its $\gamma_{usf}$ is an average of all the calculated GSF energies by considering all the possible positions of the fault planes in the SQS supercell. For example, for the 72-atom supercell, its $\gamma_{usf}$ is calculated by averaging 12 (12=6 planes $\times$ 2 directions) individual GSF energies. 

To calculate the the surface energy, $\gamma_{surf}$, a vacuum layer of 7 $\angstrom$ is inserted between two neighbouring $(1\bar10)$ planes to introduce two free surfaces, as shown in Fig. \ref{fig:SQS:cell}c. The total energy of the supercell with free surfaces is computed by only relaxing the atoms on the surface planes and their first-nearest adjacent $(1\bar10)$ planes. The rest of atoms and the supercell geometry remain fixed. The surface energy is defined as the difference in total energies of the bulk supercell (Fig. \ref{fig:SQS:cell}a) and the supercell with surfaces (Fig. \ref{fig:SQS:cell}c) dividing by twice the cross-sectional area parallel to the surface planes. Similar to $\gamma_{usf}$, $\gamma_{surf}$ of a given alloying composition is also derived by averaging the surface energies of all the $(1\bar10)$ planes in the SQS supercell. For example, $\gamma_{surf}$ of the 72-atoms supercell is calculated by averaging over six individual $(1\bar10)$ planes. 

Additionally, it is worth to mention that the convergence of the calculations on $\gamma_{usf}$ and $\gamma_{surf}$ was also tested using a series of SQSs with different sizes, as discussed in detail in Section \ref{section:verifi_comp_appro}. With the calculated $\gamma_{usf}$ and $\gamma_{surf}$, the D parameter, which qualitatively reflects the potency of the intrinsic ductility of a material\cite{rice1974ductile,rice1992dislocation,wu2015brittle,tu2019high}, can be easily derived, which is, 
\begin{equation}
\label{Eq:D_param}
    D=\gamma_{surf}/\gamma_{usf}
\end{equation}

\subsubsection{Parameter settings of DFT calculations}
In the present work, the DFT calculations were carried out using the Vienna ab-initio simulation package (VASP)\cite{Kressevasp1996}. The projector augmented wave method (PAW) \cite{BlochlPRB1994} and the exchange-correlation functional depicted by the general gradient approximation from Perdew, Burke, and Ernzerhof (GGA-PBE)\cite{Perdew1996} were employed to perform the calculations. The electronic configurations of the pseudopotentials used for the first-principles calculations are summarized in Table \ref{table:potential}. The energy cutoff of the plane-wave basis was set to be 400 eV. A first-order Methfessel–Paxton smearing of 0.2 eV was applied for Brillouin zone integration. To accommodate the differences in the size of the supercell structures, the automatic meshing scheme, as implemented in the VASP software, was used to generate the k-point grids in the first Brillouin zone of the calculations. The $R_k$ length of the automatic meshing was set to be 30\angstrom. The resulting k-point grids are $4\times2\times3$ for the 72-atom supercell in both the undeformed bulk and GSF configurations, $4\times1\times3$ for the 72-atom supercell in the surface configurations, and $4\times1\times3$ for the 90-atom supercells in either bulk or defect configurations. The energy convergence criterion of the electronic self-consistency cycle is $10^{-6}$ eV for all the calculations. For the calculations of USF and surface energies, the relaxation process is terminated when the force on each atom is less than 20 meV/Å. 

\begin{table}[H]
\renewcommand\arraystretch{1.5}
    \centering
    \caption{The electronic configurations of the pseudopotentials used in the first-principles calculations. The electrons in the bracket are treated as inner-core electrons.}
    \begin{tabular}{ccccc}
        \hline
         Element & Ti\_sv & Zr\_sv & Hf\_pv \\
         \hline
         V\_RHFIN & ([Ar])$3s^23p^63d^24s^2$ & ([Kr])$4s^24p^64d^25s^2$ & ([Xe]$4f^{14}$)$5p^65d^26s^2$ \\ 
         \hline
         Element & V\_sv & Nb\_sv & Ta\_pv \\
         \hline
         V\_RHFIN & ([Ar])$3s^23p^63d^34s^2$ & ([Kr])$4s^24p^64d^35s^2$ & ([Xe]$4f^{14}$)$5p^65d^36s^2$ \\ 
         \hline
         Element & Mo\_pv & W\_pv & Re \\
         \hline
         V\_RHFIN & ([Kr])$4p^64d^45s^2$ & ([Xe]$4f^{14}$)$5p^65d^46s^2$ &  ([Xe]$4f^{14}$)$5d^56s^2$       \\ 
         \hline
         Element & Ru &  &  \\
         \hline
         V\_RHFIN & ([Kr])$4d^65s^2$ &  &  \\ 
         \hline
    \end{tabular}
    
    \label{table:potential}
\end{table}

\subsection{Surrogate models based on statistical regression}
\label{section:surrogate_model}

Developing surrogate models for reliable predictions of USF and surface energies is necessary and crucial to enable a systematic screening of those alloy properties in a vast compositional space. In this subsection, we first describe how the descriptors for the surrogate model can be derived based on a method developed in the present work. Second, we introduce the details of a statistical regression framework used in the present work to construct the surrogate models. This framework was developed previously and has been successfully applied for modeling elastic stiffness of ordered inorganic compounds\cite{DeJong2016}.  

\subsubsection{Physics-informed descriptors}
Generating GSF and surface defects in crystals are intrinsically associated with the stretching, breaking and reforming of atomic bonds. Drawing on the idea of the bond-counting model\cite{soisson2010atomistic}, we developed a method to effectively construct a set of descriptors, which include both the information of alloy compositions and interatomic bonding characteristics. 

In an ideally mixed solid-solution alloy, atoms are not orderly organized but randomly distributed on lattice sites. The probability for two constituent elements, $i$ and $j$, to form the $i$-$j$ type of atomic bonds should equal to the product between the chemical compositions of the two elements. Based on the concept of the bond-counting model\cite{soisson2010atomistic}, we can approximate a certain physical property of an alloy, such as the cohesive energy, as a summation of the individual contributions from each atomic bond. Therefore, for the random alloys, this summation can be considered as a weighted average of the values of a physical feature, such as bond energy, associated with each type of atomic bonds. Correspondingly, the weighting factor is the presence probability of each type of atomic bonds in the alloy, which can be calculated from the alloy compositions as described above. Therefore, following the same logic, the descriptors for the surrogate models were derived as,
\begin{equation}
\label{Eq:descriptor:ave}
u_{p}=\sum_i x_i \sum_j x_j p_{ij} 
\end{equation}
Here $x_i$ and $x_j$ are the chemical compositions of element $i$ and $j$, respectively, which are among the 10 refractory elements studied in the present work (i.e., Ti, Zr, Hf, V, Nb, Ta, Mo, W, Re and Ru). $u_p$ is a descriptor developed from a bond feature parameter $p$, which has a value of $p_{ij}$ for the $i$-$j$ type of atomic bonds. Apparently, $p_{ij}$ can be written as a 10 $\times$ 10 matrix, in which each component corresponds to a pair permutation between the 10 alloying elements. In addition, to reflect the fluctuations in local atomic environments, the weighted standard deviation of $u_p$ is also considered as a descriptor($u^{\sigma}_{p}$), which is expressed as,
\begin{equation}
    \label{Eq:descriptor:dev}
u^{\sigma}_{p}=\sqrt{(\frac{1}{1-\sum_ix_i^2})\cdot(\sum_ix_i(\sum_jx_j p_{ij}-u_{p})^2)}
\end{equation}

\begin{figure}
    \centering
    \includegraphics[width=0.75\linewidth]{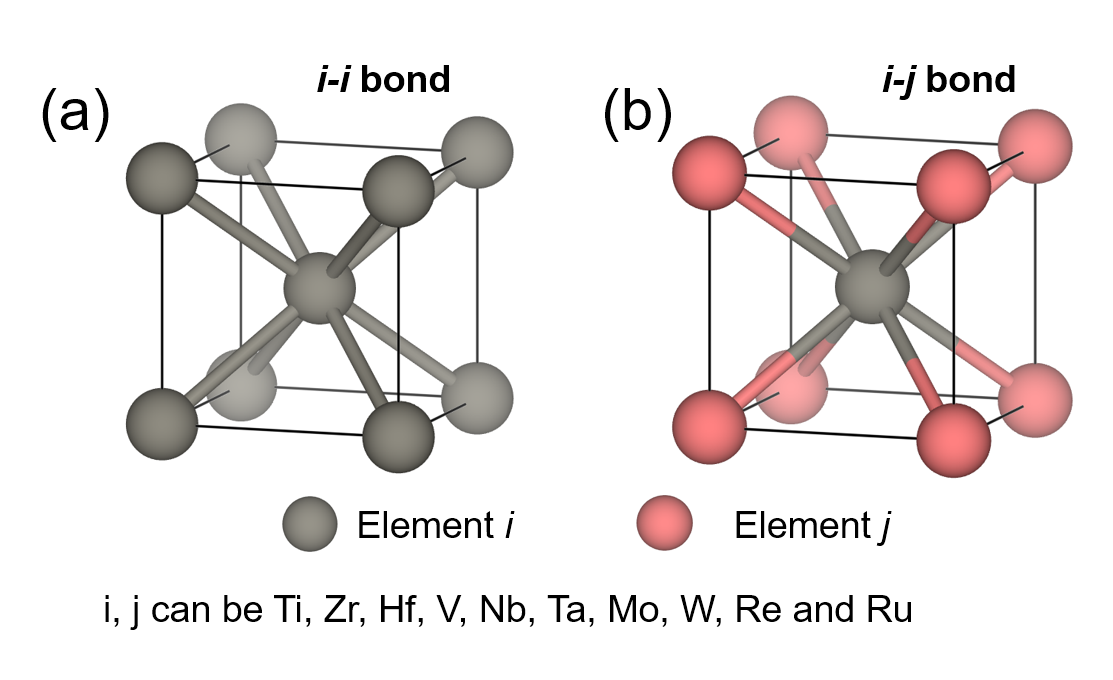}
    \caption{Schematic illustration of the first-nearest neighbor bonds in (a) the single-element bcc and (b) binary B2 structures.}
    \label{fig:bcc:cell}
\end{figure}

As a simplification, we only consider the atomic bonds in the first-nearest neighbor (FNN) shell during the bond counting process. Consequently, we could then apply the physical and electronic properties of the single-element bcc and ordered binary B2 structures as the bond feature parameters, $p_{ij}$, in Eq. \ref{Eq:descriptor:ave}. This is because, in a single-element bcc structure (e.g., composed of element $i$), the only type of the FNN atomic bond is the $i$-$i$ bond along the $\{111\}$ direction as shown in Fig. \ref{fig:bcc:cell}a. Similarly, in a B2 structure composed of element $i$ and $j$, the only type of the FNN bond is the $i$-$j$ bond shown in Fig. \ref{fig:bcc:cell}b. As an example, if we consider the cohesive energy ($E_c$) as a bond feature parameter, its $p_{ii}$ component thus equals to $E_c$ of the single-element bcc structure for the $i$-$i$ bond, while $p_{ij}$ and $p_{ji}$ both equal to $E_c$ of the binary B2 structure for the $i$-$j$ and $j$-$i$ bond, respectively.  

Since the single-element bcc and B2 structures are both ordered and highly symmetric, DFT calculations can be easily applied to generate a group of $p_{ij}$ without huge computational costs. Specifically, we performed DFT calculations to model all the possible single-element bcc and binary B2 structures constituted by the alloying elements studied in the present work. Their associated USF and surface structures were also modeled. From the DFT calculations, several physical properties of the pure elements and ordered B2 intermetallics were derived and employed as bond feature parameters, $p_{ij}$, respectively, including the USF energy($\gamma_{usf}^{bcc/b2}$), surface energy($\gamma_{surf}^{bcc/b2}$), cohesive energy ($E_c^{bcc/b2}$) and equilibrium atomic volume ($V_{eq}^{bcc/b2}$). $E_c^{bcc/b2}$ and $V_{eq}^{bcc/b2}$ were calculated by using the perfect bulk structures. $\gamma_{usf}^{bcc/b2}$ and $\gamma_{surf}^{bcc/b2}$ were calculated by using the same supercell method as described in Section \ref{section:USF_surf_method}. Before adding the shear or the vacuum layer to introduce the defects, the supercell used for the calculations for the pure metals has a geometry of $[11\bar2]\times1/2[111]\times3[1\bar10]$, while the geometry of the supercell used for the calculations of the ordered B2 intermetallics is $[11\bar2]\times[111]\times3[1\bar10]$. These supercells have the same basis vectors as the 72-atom SQSs along the $[11\bar2]$ and $[1\bar10]$ directions but shorter lengths along the $[111]$ direction due to the higher symmetries of the bcc and ordered B2 structures.

Additionally, our previous work has shown that the solute-defect interactions in refractory metals are quantitatively correlated with a group of electronic parameters\cite{DeJong2015,hu2019local}. These parameters can quantitatively describe the variations in the local electronic density of states (LDOS) of the atoms near a defect relative to those of the atoms in perfect bulk lattices. Therefore, a part of those electronic parameters were also used for the descriptor constructions. These parameters include the first and second order moments of the valance $d$- and $sp$-orbital LDOSs and the bimodality of the valance $d$-orbital LDOSs\cite{hu2019local}. The LDOSs are the projected DOSs of the atoms in the bulk lattice or on the surface or stacking fault planes of the single-element bcc metals and binary B2 alloys. The first order moment of a LDOS ($\epsilon^1_{k}$) is defined as, 
\begin{equation}
 \label{Eq:DOS}
\epsilon^1_{k}=\frac{\int_{-\infty}^{+\infty}E\rho_k(E)dE}{\int_{-\infty}^{+\infty}\rho_k(E)dE}
\end{equation}
where $\rho_k(E)$ is the DFT-calculated LDOS of the orbital $k$, and $k$ can be either the valence $d$- or $sp$- orbitals in the present work. $E$ is the band energy. Then, based on $\epsilon^1_{k}$, the second moment ($\epsilon^2_{k}$) is calculated as,
\begin{equation}
 \label{Eq:DOS_2}
\epsilon^2_{k}=\frac{\int_{-\infty}^{+\infty}(E-\epsilon^1_{k})^2\rho_k(E)dE}{\int_{-\infty}^{+\infty}\rho_k(E)dE}
\end{equation}
It should also be noted that the axis of the band energy was scaled to set the Fermi energy as zero for the integrations of Eq. \ref{Eq:DOS} and \ref{Eq:DOS_2}. Moreover, as shown in Table \ref{table:potential}, the pseudopotentials of some elements include the semi-core $s$ or $p$ electrons as valence electrons for the first-principles calculations. However, it is found that the LDOSs of these semi-core electrons localize at very low energy states and has a very large energy gap with the outermost $s$, $p$ and $d$ orbitals. We thus assume these semi-core electrons having very limited contributions to electronic bonding. Therefore, the LDOSs of these semi-core electrons are not included in the band analysis based on Eq. \ref{Eq:descriptor:ave} and \ref{Eq:descriptor:dev}. Additionally, the bimodality of a LDOS can be measured through the Hartigan’s dip test, which was described in detail in Ref \cite{hu2019local}. 

Moreover, two features of elemental properties, the numbers of valance electrons and the Pauling electronegativity, are also used to construct the descriptors in the present work. Specifically, the descriptors were calculated as, 
\begin{equation}
    \label{Eq:descriptor:ele_ave}
u_{q}=\sum_ix_iq_i
\end{equation}

\begin{equation}
    \label{Eq:descriptor:ele_dev}
u_{q}^{\sigma}=\sqrt{(\frac{1}{1-\sum_ix_i^2})\cdot(\sum_ix_i(q_{i}-u_{q})^2)}
\end{equation}
where $q_i$ represents either the numbers of valance electrons or the Pauling electronegativity of element $i$ and $x_i$ represents the chemical composition of $i$. The descriptors generated by Eq. \ref{Eq:descriptor:ele_ave} and \ref{Eq:descriptor:ele_dev} are intended to describe the average filling level of the d-bands and the tendency of charge transfer between different elements, respectively. 

Table \ref{Table:feature} lists all the atomic bond parameters ($p_{ij}$) and elemental properties ($q_i$) used for descriptors construction in the present work. Each of them results in two descriptors: one relates to the arithmetic mean (Eq. \ref{Eq:descriptor:ave} and \ref{Eq:descriptor:ele_ave}) and the other one relates to the weighted standard deviation of the mean (Eq. \ref{Eq:descriptor:dev} and \ref{Eq:descriptor:ele_dev}). Therefore, 42 descriptors were generated in total. Noteworthily, only 30 of 42 descriptors were employed as input variables for the regression of $\gamma_{usf}$, where the descriptors associated with the surface structures were not used. In the same way, when the regression was performed for $\gamma_{surf}$, the descriptors associated with the stacking faults were not used.

\subsubsection{Statistical regression framework}

In the present work, a statistical regression framework\cite{DeJong2016}, namely Gradient Boosting Machine Local Polynomial Regression (GBM-Locfit), was used to perform the regression analysis and obtain quantitative models to predict $\gamma_{usf}$ and $\gamma_{surf}$ of bcc solid-solution alloys. In the GBM-Locfit framework, the gradient boosting machine iteratively produces a prediction model in the form of an ensemble of predictive functions, $\eta_i$. At each GBM-iterative step $i$, $\eta_i$ is generated from a regression model implemented in the Locfit package\cite{loader2006local}, which performs kernel-based multivariate locally linear regressions. In order to reduce the risk of over-fitting, at each GBM step, only a subset of the input descriptors were used for performing the Locfit regression. Specifically, many individual regressions were performed to traverse all the possible subsets of the descriptor sets generated from Table \ref{Table:feature}. Then, only the predictive function from the regression that leads to the greatest reduction in the residue from the previous GBM steps was added as a $\eta_i$ into the ensemble of predictive functions. For the regression of $\gamma_{usf}$, the size of the subset is set to be not exceed three descriptors, while this limit is set to be 2 for the regression of $\gamma_{surf}$. Therefore, not all the input descriptors were used during the whole regression process, and only the most relevant ones were selected automatically by the GBM-Locfit framework\cite{DeJong2016}. The final prediction model was derived as the sum of all the predictive functions, while each function was attenuated by a learning rate of 0.05. Moreover, the GBM-locfit framework is performed with n-fold cross-validation (n depends numbers of regression samples) and a conservative risk criterion to determine a optimal number of iteration steps\cite{DeJong2016,hu2020predicting}. 

Due to the high computational cost, the DFT calculations with the SQS method were only able to generate the $\gamma_{usf}$ and $\gamma_{surf}$ data for a finite amount of bcc solid-solution alloys (Table \ref{SQS:list}). Therefore, instead of randomly splitting the DFT data into regression and test sets, we specifically trained the model with the data of binary and ternary alloys only, and used the rest quaternary data as the test set, which allows us to maximally validate the predictive ability of the generated surrogate model for multicomponent alloy compositions. After this validation, all the data listed in Table \ref{SQS:list} were used as the training set to update the models for screening purposes. The screening predictions from the newly updated model were further validated by an additional set of DFT calculations as described in Fig. \ref{fig:screening_validation}.

\section{Results and Discussion}
\subsection{Verification of the computational approach for $\gamma_{usf}$ and $\gamma_{surf}$}
\label{section:verifi_comp_appro}
By definition, the unstable stacking fault energy, $\gamma_{usf}$, corresponds to the maximum value of the GSF curve, which is generally obtained by interpolating the results of a series of GSF calculations performed at different shift distances. This will severely increase the computational cost if the target is to investigate $\gamma_{usf}$ for a large amount of alloying compositions, for example, 106 compositions in the present work. On the other hand, due to the mirror symmetry of the $(1\bar10)$ plane, $\gamma_{usf}$ of the $(1\bar10)[111]$ slip in a bcc lattice should occur at a shift distance of $|\frac{1}{4}[111]|$ although deviations may be induced by the local chemical variations and lattice distortions\cite{xu2020atomistic}. In the present work, for the sake of simplification, we used the GSF energies that correspond to the $|\frac{1}{4}[111]|$ shifts on different fault planes in the supercell to derive the averaged $\gamma_{usf}$ for a given alloy composition. Therefore, benchmark calculations are necessary to verify such simplification. 

\begin{figure}[htbp]
    \centering
    \includegraphics[width=1.0\linewidth]{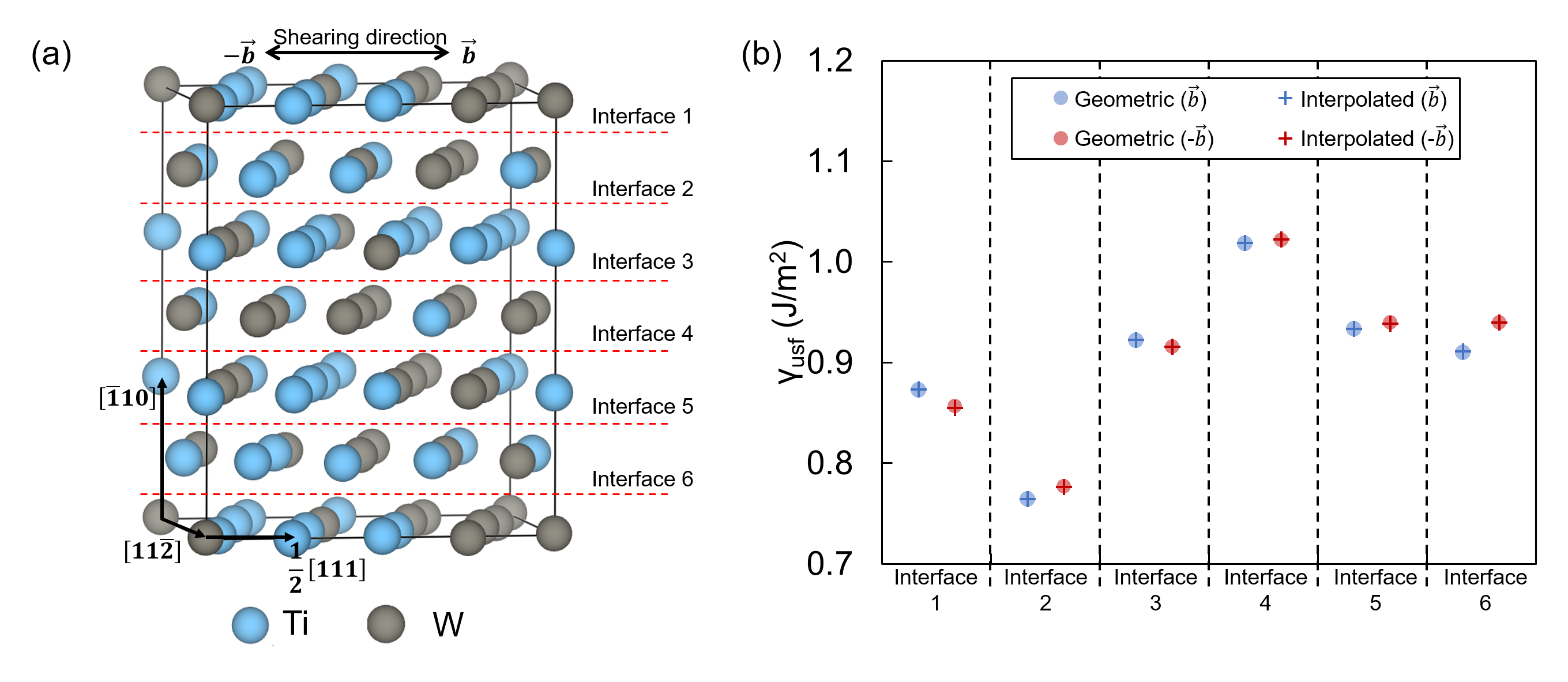}
    \caption{(a) The supercell structure for the equimolar TiW alloy after relaxation. To obtain the final averaged $\gamma_{usf}$ of the supercell, calculations of GSF energies have to be performed for six individual interfaces, and each interface has two shift directions ($-\vec{b}$ and $\vec{b}$). (b) The geometric $\gamma_{usf}$ (circle symbols) of different interfaces marked in Fig. \ref{fig:gsf_benchmark}a in comparison with the corresponding interpolated $\gamma_{usf}$ (cross symbols). The data points corresponding to the shift along the $\vec{b}$ direction are marked in blue color, while those associated with the opposite shift direction are marked in red color. The geometric $\gamma_{usf}$ corresponds to the GSF energy at a shift of $|\frac{1}{4}[111]|$. The interpolated $\gamma_{usf}$ is derived by interpolating the maximum point of the GSF curve.} 
    \label{fig:gsf_benchmark}
\end{figure}

Here, using the equimolar TiW alloy as an example, benchmark calculations were performed to investigate the difference between the GSF energy at a shift of $|\frac{1}{4}[111]|$ (referred as the geometric $\gamma_{usf}$ in the following) and the maximum value interpolated from the GSF curve (referred as the interpolated $\gamma_{usf}$ in the following). As shown in Fig. \ref{fig:gsf_benchmark}a, the relaxed supercell structure of the TiW alloy has six individual interfaces between the neighboring $(1\bar10)$ planes to generate stacking faults. Each of these interfaces should have two sets of GSF energies corresponding to the shifts along the $-\vec{b}$ and $\vec{b}$ directions, respectively, where $\vec{b}$ is the Burgers vector $\frac{1}{2}[111]$. For a given shift direction of a specific interface, we first calculated the GSF energies at a shift of $0.5|\vec{b}|$, which yield the values of the geometric $\gamma_{usf}$. Then, additional calculations were performed at shifts of $0.375|\vec{b}|$, $0.4375|\vec{b}|$, $0.5625|\vec{b}|$, and $0.625|\vec{b}|$, respectively. Together with the GSF energies at $0.5|\vec{b}|$, a second-order polynomial fitting was performed for the five data points with respect to their shift distances. The maximum of the fitted polynomial yields the value of the interpolated $\gamma_{usf}$. The $R^2$ of the fittings for all the interfaces were found to be close to 1 ($>$0.99), which means that the GSF curve around its maximum can be well described by a second-order polynomial. Fig. \ref{fig:gsf_benchmark}b shows a comparison between the values of the geometric and interpolated $\gamma_{usf}$ for each of the possible interfaces and shift directions in the supercell of Fig. \ref{fig:gsf_benchmark}a. These results suggest the energy differences between the geometric and interpolated $\gamma_{usf}$ are almost negligible compared to the energy difference between different stacking fault interfaces. Since the final $\gamma_{usf}$ of the supercell is derived by averaging the results over all the interfaces along both shift directions, using the GSF energy at a shift of $|\frac{1}{4}[111]|$ for the derivation will not notably impact the result accuracy. In the meanwhile, the computational efficiency is significantly improved. 

\begin{figure}[htbp]
    \centering
\subfigure{
    \centering
    \includegraphics[width=0.9\linewidth]{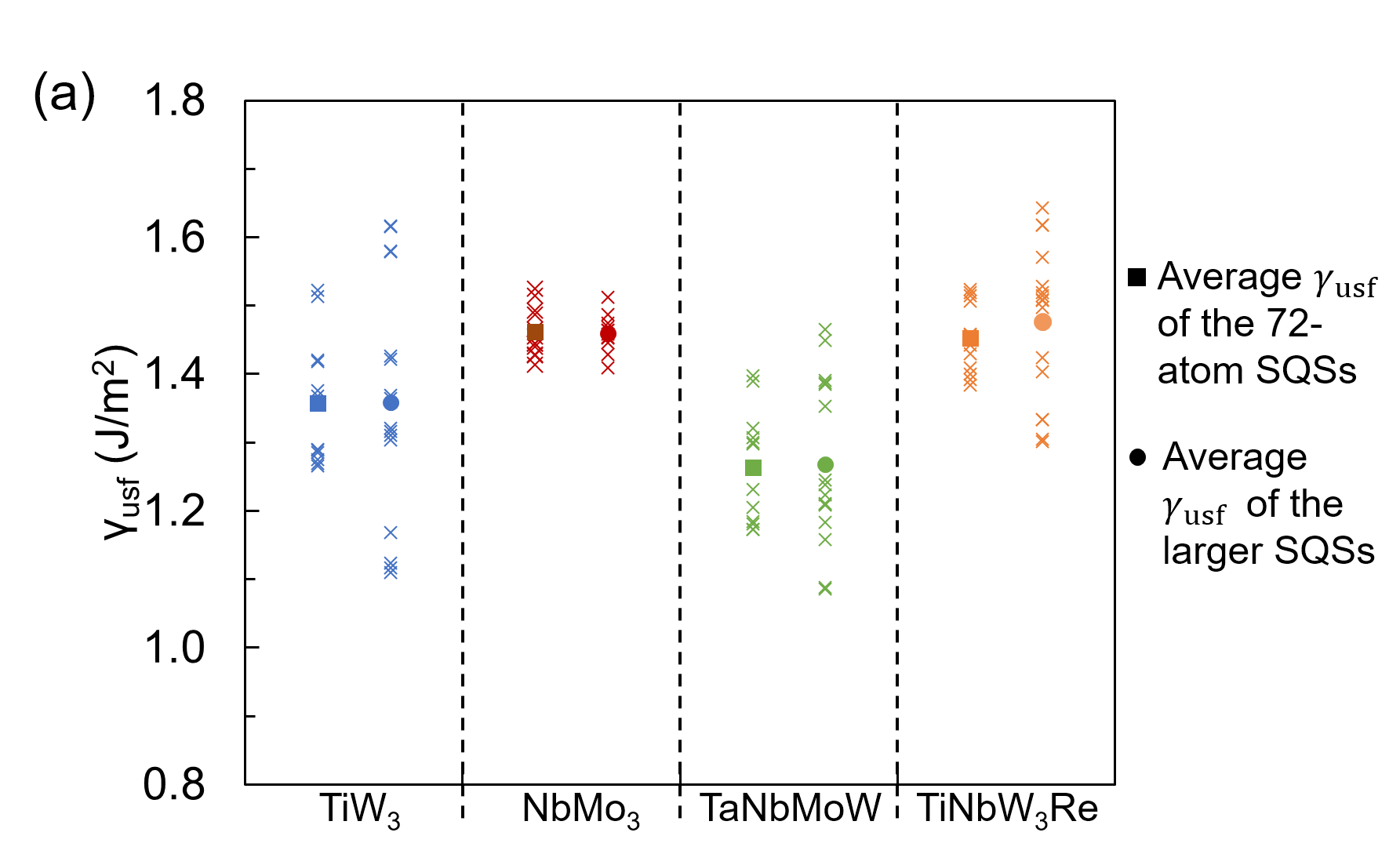}}
\subfigure{
    \centering
    \includegraphics[width=0.9\linewidth]{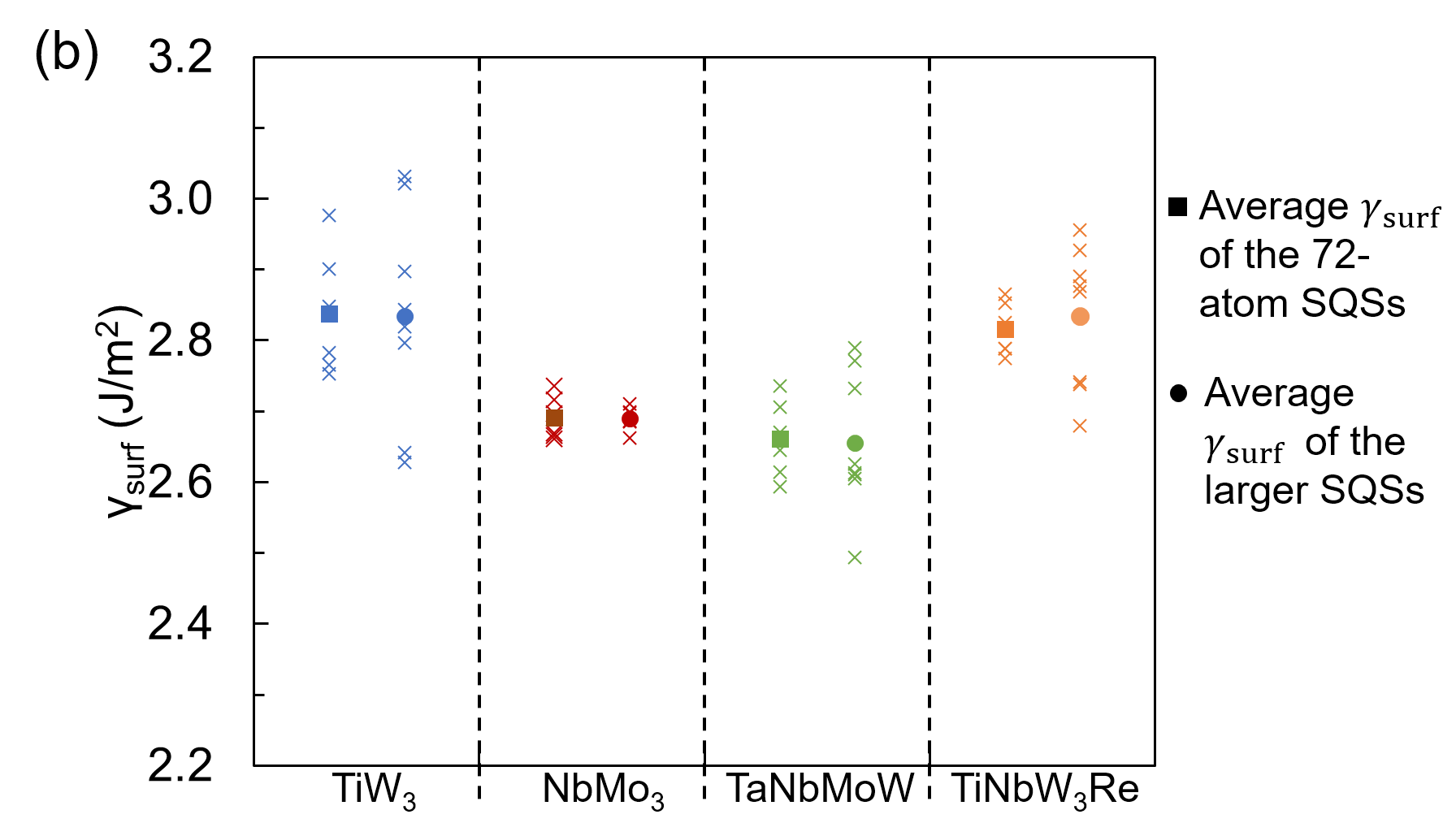}}
    \caption{(a) Unstable stacking fault energies and (b) surface energies of the TiW$_3$, NbMo$_3$, TaNbWMo and TiNbW$_3$Re alloys calculated using the SQS supercells with different sizes. For each of the alloy compositions, the solid cubic or circle symbol corresponds to the averaged $\gamma_{usf}$ and $\gamma_{surf}$ calculated using the original 72-atoms supercell or a larger supercell, respectively. The cross symbols correspond to the individual values of $\gamma_{usf}$ and $\gamma_{surf}$ due to the different choices on the positions of the defect planes in the supercell.} 
    \label{fig:size_benchmark}
\end{figure}

Furthermore, benchmark calculations were also performed to evaluate the convergence of the calculated $\gamma_{usf}$ and $\gamma_{surf}$ with respect to the size of the SQS supercells. Using the ATAT code\cite{VanDeWalle2013}, four individual SQS supercells with sizes larger than the original 72-atom supercell were additionally generated for the alloy compositions of TiW$_3$, NbMo$_3$, TaNbWMo and TiNbW$_3$Re, respectively. Specifically, the basis vectors of these additionally generated SQS supercells are $[11\bar2]\times2[111]\times4[1\bar10]$, $2[11\bar2]\times2[111]\times3[1\bar10]$, $[11\bar2]\times2[111]\times4[1\bar10]$, and $[11\bar2]\times3[111]\times4[1\bar10]$, respectively. As a comparison, the basis vectors of the original 72-atom supercell are $[11\bar2]\times2[111]\times3[1\bar10]$. Then, following the same method described in Section 2.1.1, we calculated $\gamma_{usf}$ and $\gamma_{surf}$ using these larger supercells, and compare the results with those using the original 72-atom supercell. As shown in Fig. \ref{fig:size_benchmark}, increasing the supercell size has very limited effects on the averaged $\gamma_{usf}$ and $\gamma_{surf}$, even though it could result in larger deviations among the individual USF and surface energies that correspond to the different choices on the positions of the defect planes in the supercell.

\subsection{DFT results on $\gamma_{usf}$ and $\gamma_{surf}$}
\begin{figure}[htbp]
    \centering
    \includegraphics[width=1.0\linewidth]{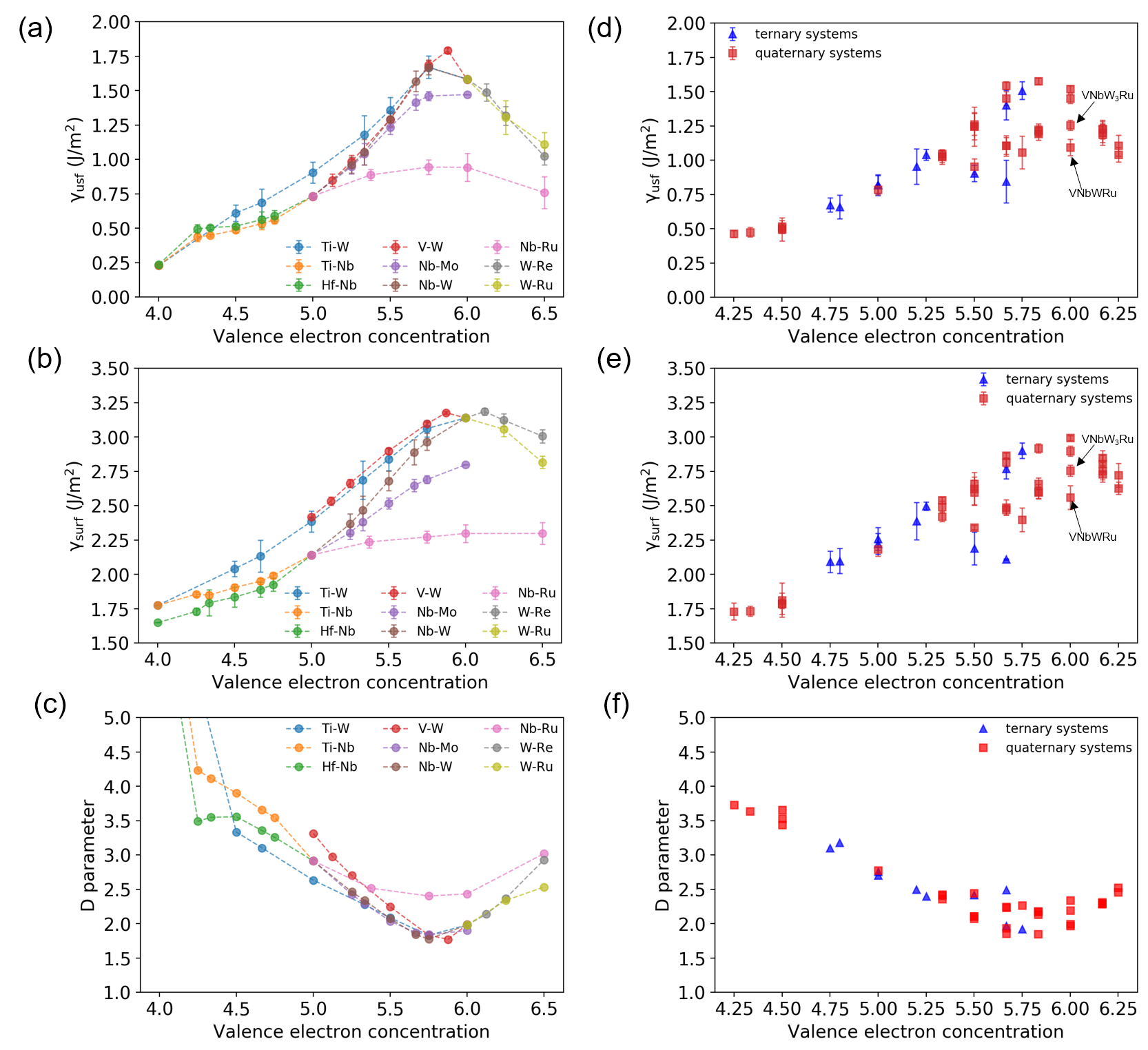}
    \caption{The unstable stacking fault energy ($\gamma_{usf}$), surface energy ($\gamma_{surf}$) and $D$ parameter of the bcc refractory alloys predicted from DFT calculations. (a) $\gamma_{usf}$, (b) $\gamma_{surf}$, and (c) $D$ parameter of the binary alloys plotted with respect to the valence electron concentration (VEC) of the alloys. The alloys from the same binary system are marked by the same color. Each of the binary systems corresponds to an individual color in the legend. (d) $\gamma_{usf}$, (e) $\gamma_{surf}$, and (f) $D$ parameter of the ternary (blue triangles) and quaternary alloys (red squares) plotted with respect to the VEC of the alloys. It should be noted that each of the data points in (a), (b), (d) and (e) corresponds to an average value of $\gamma_{usf}$ or $\gamma_{surf}$  for an alloy composition, which is obtained by taking an average over all the possible locations of the defect planes in the SQS supercell. The error bar corresponds to the standard deviation of the average.} 
    \label{fig:DFTvsVEC}
\end{figure}

In the following, we further discuss the correlations of these results with the filling fraction of the valence $d$ orbitals in Fig. \ref{fig:DFTvsVEC}, as the d-band filling effect is generally essential in determining  physical and mechanical properties of the transition metal elements and their alloys\cite{qi2014tuning,hu2019local,hu2017solute,al2019elastic}.

 $\gamma_{usf}$ and $\gamma_{surf}$ of various binary alloys are plotted with respect to the alloy's valence electron concentration (VEC) in Fig. \ref{fig:DFTvsVEC}a and \ref{fig:DFTvsVEC}b, respectively. The VEC of an alloy is calculated as the average over the number of valence electrons of the constituent elements with respect to their mole fractions (unit: $e^-$/atom). In transition metal alloys, a higher value of VEC corresponds to a higher average filling fraction of the valence $d$-bands. In the present work, the valence electrons of an element are considered as its outermost $s$ and $d$ electrons. Specifically, for the $3d$ elements, only the $3d$ and $4s$ electrons are considered as valence electrons. The same definition of valence electrons are used for the $4d$ and $5d$ elements. The binary alloys in Fig. \ref{fig:DFTvsVEC}a and \ref{fig:DFTvsVEC}b are selected from nine different alloy systems as they are representative to show clear trends of the USF and surface energies with respect to VEC. Clearly, both $\gamma_{usf}$ and $\gamma_{surf}$ of the binary alloys show a nearly parabolic dependence on the variations of VEC. In a range of VEC from 4.0 to 6.5 $e^-$/atom, $\gamma_{usf}$ and $\gamma_{surf}$ first increase with VEC to reach a maximum, and then decrease with the further increases of VEC. It is interesting to note that the highest $\gamma_{usf}$ does not correspond to pure bcc W but the VW$_7$ alloy with a VEC of 5.875 $e^-$/atom. This indicates that the mechanical strength of the group VI metals can be further enhanced by properly alloying them with a small amount of group IV or V metals. However, over alloying of group IV or V metals could oppositely trig a softening effect due to the parabolic behavior of $\gamma_{usf}$. Similar results were also reported for the ideal tensile behavior of binary refractory alloys\cite{qi2014tuning,yang2018ab}. 
 
 Moreover, it is noted that the quantitative dependences of $\gamma_{usf}$ and $\gamma_{surf}$ on VEC are very different between individual binary systems. In other words, for different alloy compositions with the same VEC, their $\gamma_{usf}$ and $\gamma_{surf}$ can still vary significantly. For example, as shown in Fig. \ref{fig:DFTvsVEC}a, $\gamma_{usf}$ of the NbMo$_3$ alloy is calculated to be about 50\% higher than that of the Nb$_3$Ru alloy, even though both two alloys have the VEC of 5.75 $e^-$/atom and are composed of the elements in the same period. The large differences in USF and surface energies are also observed among the binary alloys with identical VECs in the Ti-W, Ti-Nb, V-W and Nb-W systems. Additionally, as shown in Fig. \ref{fig:DFTvsVEC}c, the $D$ parameter of the binary alloys also qualitatively shows an approximately parabolic function of VEC, but in an inverse manner to that of $\gamma_{surf}$ and $\gamma_{usf}$. Expect for the Nb-Ru system, the curves of other alloy systems are almost overlapped in the range of VEC from 5.25 to 6.25 $e^-$/atom, where the alloys also generally have a large $\gamma_{usf}$. The results suggest that a binary refractory alloy that corresponds to a higher mechanical strength would generally have a poorer ductility, vice versa. Therefore, it is difficult to simultaneously improve both the mechanical strength and ductility of binary refractory alloys by tuning their chemical compositions. 

Interestingly, it is found that the correlations of the USF and surface energies with VEC in multicomponent refractory alloys is ambiguous and much weaker compared to that of the binary alloys. As shown in Fig. \ref{fig:DFTvsVEC}d and \ref{fig:DFTvsVEC}e, $\gamma_{usf}$ and $\gamma_{surf}$ of the ternary and quaternary alloys listed in Table \ref{SQS:list} are also plotted against their VEC values, respectively. Clearly, there are multiple alloys that have the same VEC but very different USF and surface energies compared with each other. Even for the alloys with the same constituent elements, their $\gamma_{usf}$ and $\gamma_{surf}$ are still not solely determined by VEC, such as the VNbWRu and VNbW$_3$Ru alloys marked by arrows in Fig. \ref{fig:DFTvsVEC}d and \ref{fig:DFTvsVEC}e. These results suggest that simply comparing the VEC values between two multicomponent alloys may not be able to even qualitatively distinguish the difference in their USF and surface energies. As shown in Fig. \ref{fig:DFTvsVEC}f, the $D$ parameter of the multicomponent alloys are also distributed in a more scattered pattern on VEC compared to that of the binary alloys.These results are expected because the original $d$-band filling effects on alloy properties, such as cohesive energies, were derived using the classical Friedel model\cite{sutton1993electronic} by assuming the DOS of $d$ orbitals has a fixed rectangular shape, which is not accurate in realistic alloys\cite{hu2019local}. Therefore, in order to quantitatively predict the USF and surface energies of the bcc refractory alloys, additional physics-informed descriptors are required to describe the unevenly distributed chemical bonds and electronic structures in the alloys. The corresponding attempts were made in the present work based on a bond-counting model and discussed in details in the following section. 

\subsection{Predictive ability of the surrogate model}

Although the DFT calculations with the SQS method provide means to predict the USF and surface energies for the ideally mixed solid-solution alloys, it is still practically infeasible to directly apply it for screening a vast compositional space due to the extensive computational cost. Therefore, in the present work, surrogate models were developed using a statistical regression framework, GBM-locfit\cite{DeJong2016}, to efficiently predict the USF and surface energies of bcc refractory alloys in a large multicomponent compositional space containing 10 different alloying elements (i.e., Ti, Zr, Hf, V, Nb, Ta, Mo, W, Re and Ru). As described in Section \ref{section:surrogate_model}, the surrogate models rely on a set of physics-informed descriptors generated from a bond-counting approach and the training data generated from the DFT calculations. 

\begin{figure}[htbp]
    \centering
    \includegraphics[width=1.0\linewidth]{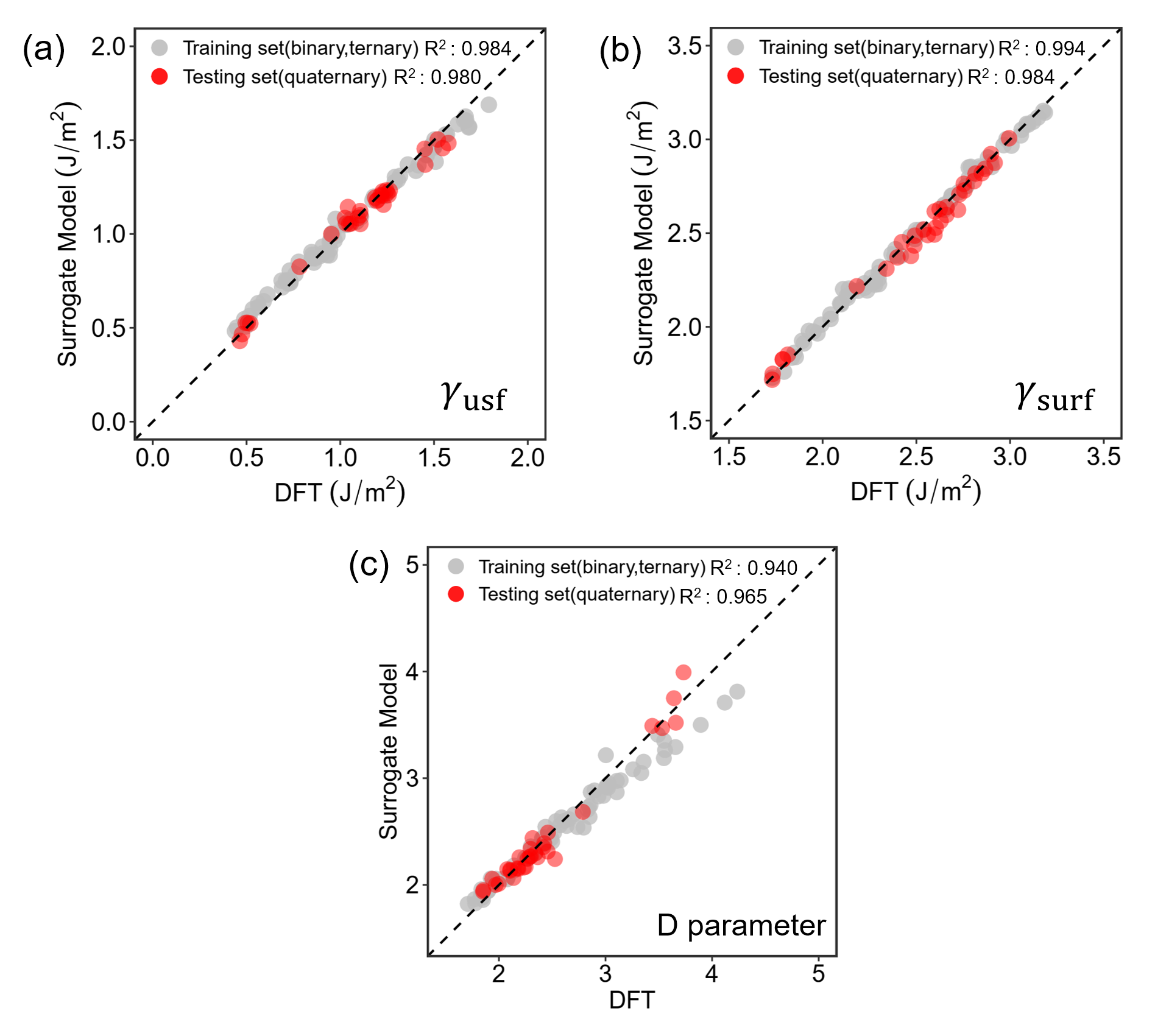}
    \caption{ Training and testing performance of the developed surrogate models for (a) unstable stacking fault energy ($\gamma_{usf}$), (b) surface energy ($\gamma_{surf}$), and (c) $D$ parameter. The training set is only composed of the binary and ternary alloys in Table \ref{SQS:list}. The test set is composed of the quaternary alloys in Table \ref{SQS:list}. The data of the training and test set are marked in gray and red color, respectively. Additionally, it should be noted that the $D$ parameter was never used as a regression response to train the surrogate model. The predictions of the surrogate model in (c) are derived from the corresponding predictions on $\gamma_{usf}$ and $\gamma_{surf}$ based on Eq. \ref{Eq:D_param}.}

    \label{fig:ML_binay_quat}
\end{figure}

To evaluate the prediction capability of the surrogate models, especially for multicomponent alloy systems, we specifically trained the model only with the DFT data of the binary and ternary alloys listed in Table \ref{SQS:list} (74 individual alloy compositions in total), and employed the rest quaternary data (32 individual alloy compositions in total) as a test set never used for training. The training and testing results of the surrogate model on $\gamma_{usf}$ are presented in Fig. \ref{fig:ML_binay_quat}a, where the model predictions are plotted against the results of DFT calculations. As shown by the grey dots in Fig. \ref{fig:ML_binay_quat}a, $\gamma_{usf}$ of the alloys in the training set are well reproduced by the surrogate model, yielding an root-mean-squared error (RMSE) of 0.047 $J/m^2$ and a $R^2$ value about 0.984. Moreover, as shown by the red dots in Fig. \ref{fig:ML_binay_quat}a, by only trained with the binary and ternary data, the model can provide reliable predictions for the quaternary alloys in the test set. The RMSE of the model predictions on the test set is only about 0.043 $J/m^2$, close to that of the training set, and the corresponding $R^2$ value for testing is 0.980. 

In the same way, the training and testing performance of the surrogate model on $\gamma_{surf}$ is illustrated in Fig. \ref{fig:ML_binay_quat}b. It also is shown that the model trained only based on the binary and ternary data is able to accurately predict $\gamma_{surf}$ of the quaternary alloys in the test set. The RMSE of the predictions on $\gamma_{surf}$ is 0.030 and 0.046 $J/m^2$ for the training and test set, respectively, while the corresponding $R^2$ value is 0.994 and 0.984, respectively. Furthermore, it is worth noting that none of the alloys in the training set includes Zr as their constituent elements but the models still accurately predict $\gamma_{usf}$ and $\gamma_{surf}$ for the Zr-containing alloys in the test set. The results of Fig. \ref{fig:ML_binay_quat}a and \ref{fig:ML_binay_quat}b support that the developed surrogate models can efficiently and effectively predict the variations of the USF and surface energies in a large compositional space for the bcc refractory alloys. 

Additionally, with the predictions of the surrogate models on $\gamma_{usf}$ and $\gamma_{surf}$, the $D$ parameter of the alloys in both training and test sets can be easily derived based on Eq \ref{Eq:D_param}. As shown in Fig. \ref{fig:ML_binay_quat}c, the values of the $D$ parameter derived from the model predictions are generally in good agreement with the results obtained from the DFT calculations. In addition, for the alloys in the training set with relatively larger $D$ parameters, the results from the surrogate models seem to be systematically lower than the DFT values. This is caused by the slight overestimation of the surrogate model on $\gamma_{usf}$ for the alloys with relatively low USF energies (Fig. \ref{fig:ML_binay_quat}a), which is not seen for the predictions of the surface energy (Fig. \ref{fig:ML_binay_quat}b). This discrepancy should have limited effects on the final results of screening for alloy compositions with enhanced strength-ductility synergies, because the relative disparity of the $D$ parameter between different alloys are still generally captured by the surrogate models.  

\subsection{Screening of the alloy properties in multicomponent systems}
\label{section:screening}

Because of our descriptor construction method, the developed surrogate models are able to make immediate predictions on $\gamma_{usf}$, $\gamma_{surf}$ and the $D$ parameter by only requiring the information of alloy compositions, without the need of any additional DFT calculations. Therefore, the models are quite suitable to perform rapid screenings of these alloy properties in complex compositional spaces. In the present work, we applied the developed surrogate models to systematically screen $\gamma_{usf}$ and $\gamma_{surf}$ of bcc refractory solid-solution alloys in a multicomponent compositional space to search promising alloy compositions with enhanced strength-ductility synergy. The compositional space chosen for screening is composed of 10 different transition metal elements, which are Ti, Zr, Hf, V, Nb, Ta, Mo, W, Re and Ru. The first 8 elements are commonly included as constituent elements in bcc refractory HEA and multicomponent alloys\cite{senkov2018development,gao2016design}. Re and Ru were also included because these two elements, especially Re, were recognized to improve the low-temperature ductility of Group VI bcc metals under proper alloying amounts\cite{ren2018methods,klopp1968review}. In addition, several recent works also reported successful syntheses of novel bcc refractory HEAs that containing Re as one of the principle elements\cite{gao2016design,tong2020severe,zhang2020microstructure}. Therefore, including Re and Ru into the screening space would further expand our theoretical search to cover more unexplored and unconventional alloy compositions in which promising candidates may exist. 

Specifically, the screenings were performed over all the quaternary alloy systems in the 10-component space. For each of the quaternary systems, its compositional space was evenly grided using an interval of $\frac{1}{18}$ mole fraction along each axis that stands for an elemental concentration. The predictions on $\gamma_{usf}$ and $\gamma_{surf}$ were then performed at individual alloy compositions that correspond to the grid points, and the value of the $D$ parameter was correspondingly derived based on Eq. \ref{Eq:D_param}. Moreover, during the screening, the mole fraction of Re and Ru are constrained to be no more than 0.25 and 0.08333, respectively, by considering their limited binary solubility in the bcc phase of the group V and VI metals\cite{osti_4051537,mathieu2013calphad,liu2013first}. The reason that we only screened over all the quaternary systems by a discrete compositional interval is for the convenience of further validations by DFT calculations using SQS supercells. It is known that the necessary supercell size for generating a reliable SQS increases drastically with the number of constituent elements of the system. By applying such screening constraints, any of the screened alloy compositions can be easily accessed by generating a 72-atom SQS supercell. This would allow us to validate the screening results of many different alloy compositions under affordable computational costs. Furthermore, it is worth mentioning that the actual prediction range of the surrogate models is not limited to the quaternary alloys. The models can efficiently predict $\gamma_{usf}$ and $\gamma_{surf}$ for any multicomponent alloys with continuous compositional variations in the 10-element compositional space. 

It has to be emphasized that, to possibly improve the accuracy of the screening results, the surrogate models were further re-trained with all the data in Table \ref{SQS:list}, instead of splitting the data into training and test sets, and then applied for the screening calculations. The newly trained models would possibly yield more reliable predictions compared with those used for Fig. \ref{fig:ML_binay_quat} because the new models included more information on complex multicomponent alloy systems during the training process. Since all the data in Table \ref{SQS:list} were employed for training, the predictions of the newly trained models were validated by performing additional DFT calculations with the SQS method as discussed above. 

\begin{figure}[H]
    \centering
    \includegraphics[width=1.0\linewidth]{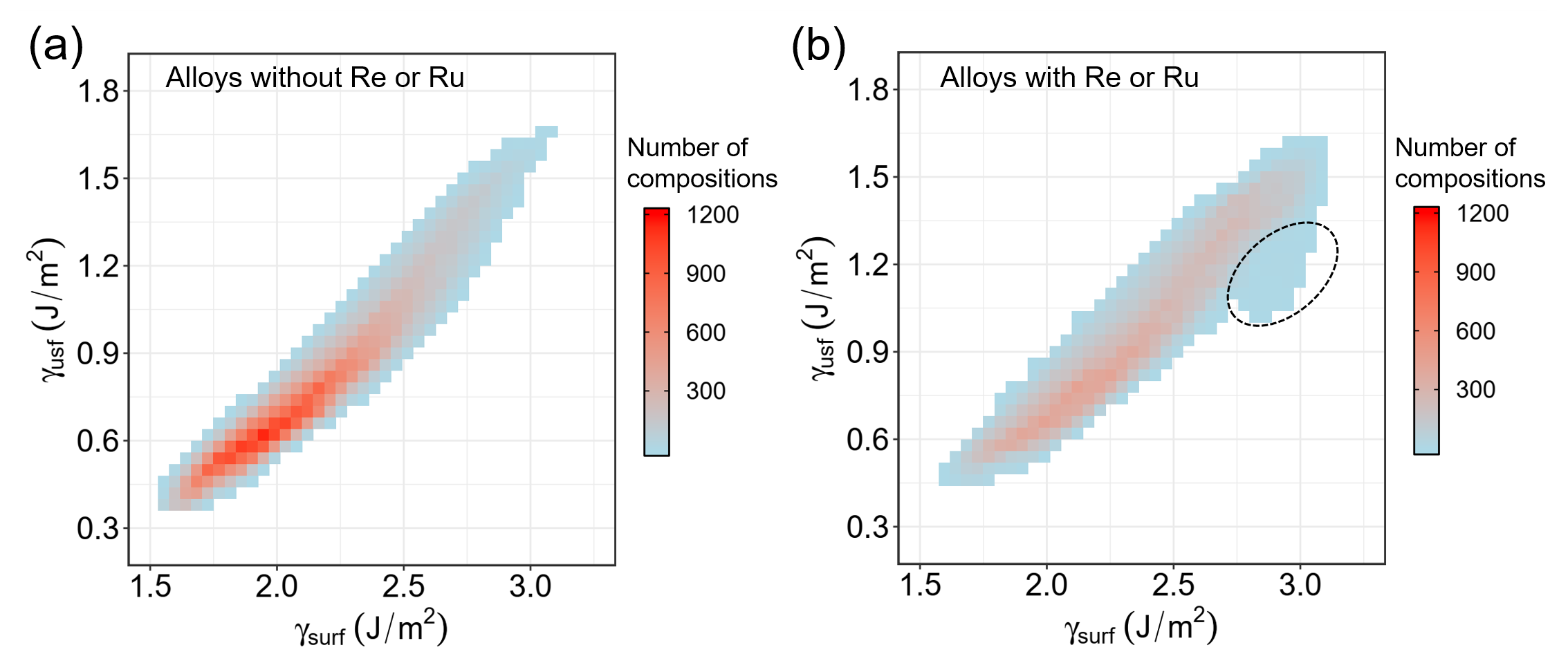}
    \caption{Distribution of the USF ($\gamma_{usf}$) and surface ($\gamma_{surf}$) energies predicted by the newly trained surrogate models through the screening of 112,378 quaternary alloys in the 10-component compositional space. (a) Results of the alloys without Re or Ru as constituent elements; (b) Results of the alloys with Re or Ru as constituent elements. A squared unit with a warmer color means that there are more alloy compositions having $\gamma_{usf}$ and $\gamma_{surf}$ within the coverage area of the unit. The dashed circle marks a unique distribution pattern of the USF and surface energies from a part of the Re/Ru-containing alloys. The alloys corresponded to the pattern may have strong potency of being intrinsic ductile and maintain relatively high mechanical strengths.}
    \label{fig:usf_surf_distri}
\end{figure}

Overall, the newly trained surrogate models have been applied to predict $\gamma_{usf}$, $\gamma_{surf}$ and $D$ parameters for 112,378 alloy compositions in 210 different quaternary systems. In order to distinguish the effects of Re and Ru, the prediction results are grouped into two sets for visualization. One of the sets corresponds to the alloys containing Re or Ru, while the other set corresponds to the rest alloys with Re/Ru-free compositions. The distributions of the USF and surface energies from the screening predictions are illustrated in Fig. \ref{fig:usf_surf_distri} using 2D density plots, in which a squared unit with a warmer color means that there are more alloy compositions having $\gamma_{usf}$ and $\gamma_{surf}$ within the coverage area of the unit. As shown in Fig. \ref{fig:usf_surf_distri}a, in the alloys without Re or Ru, the variations of $\gamma_{usf}$ overall exhibits a positive correlation with that of $\gamma_{surf}$. On the other hand, as shown in Fig. \ref{fig:usf_surf_distri}b, introducing Re or Ru as alloying element leads to a more spread distribution between $\gamma_{usf}$ and $\gamma_{surf}$. Particularly, as indicated by the dashed circle in Fig. \ref{fig:usf_surf_distri}b, the USF and surface energies of a part of Re/Ru-containing alloys show a unique distribution pattern, which is not seen in the results of the Re/Ru-free alloys (Fig. \ref{fig:usf_surf_distri}a). The alloys corresponding to this pattern generally have surface energies close to the maximum of the screening results but maintain moderate USF energies around 1.2 $J/m^2$, consequently yielding larger $\frac{\gamma_{surf}}{\gamma_{usf}}$ ratios relative to the Re/Ru-free alloys with the same level of USF energies. 

\begin{figure}[H]
    \centering
    \includegraphics[width=1.0\linewidth]{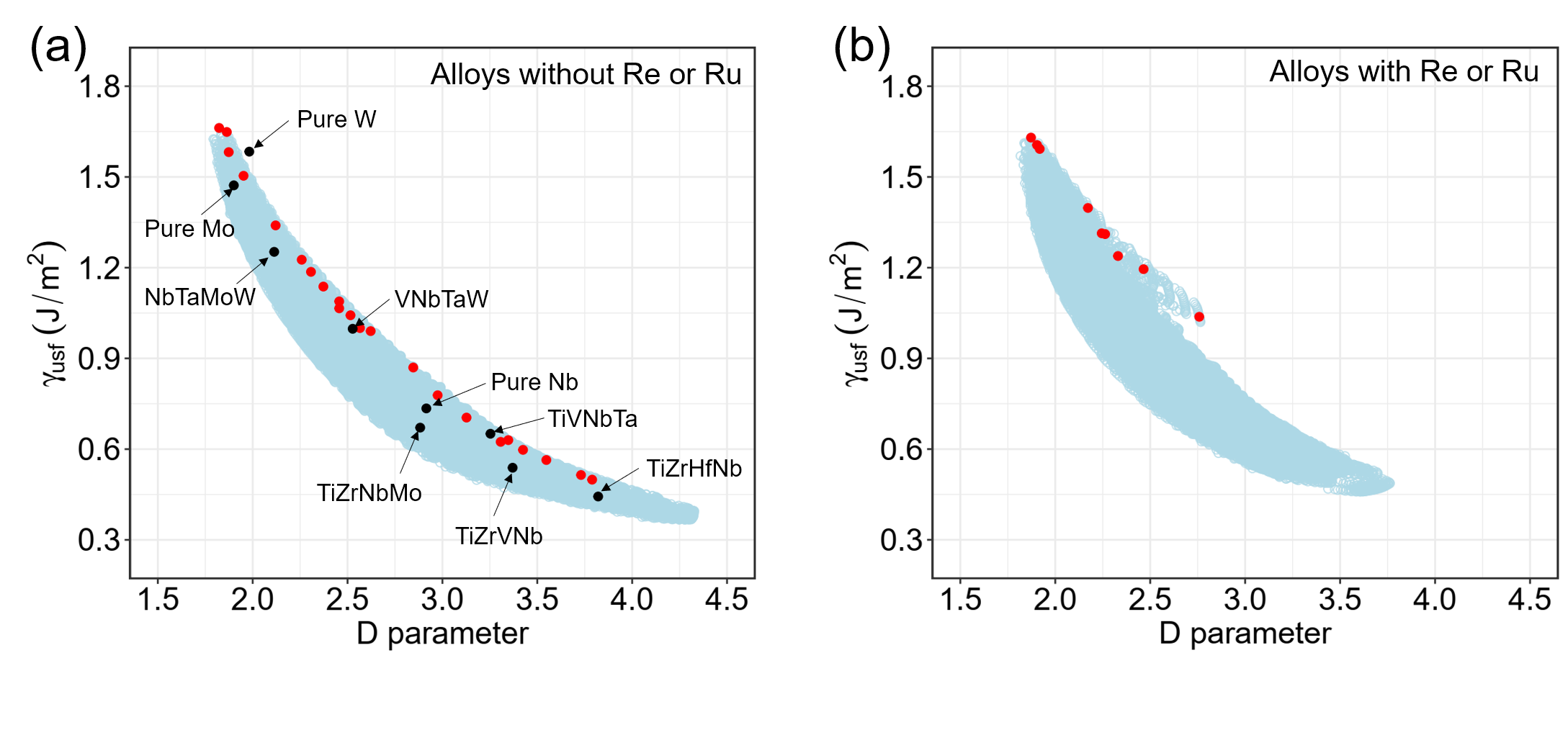}
    \caption{Plots of $\gamma_{usf}$ vs. the $D$ parameter obtained by screening over 112,378 quaternary alloys in the 10-component compositional space. (a) Results of the alloys without Re or Ru as constituent elements; (b) Results of the alloys with Re or Ru as constituent elements. The screening results are represented by the open circles with light-blue color. As a comparison, we also particularly marked the prediction results of some known bcc refractory HEAs and pure bcc metals using solid circles with black color. The data points additionally marked in red correspond to those selected for the DFT validations as shown in Fig. \ref{fig:screening_validation}.}

    \label{fig:usf_D_distri}
\end{figure}

The screening results are also visualized by plotting the variations of $\gamma_{usf}$ with respect to the $D$ parameter, as shown in Fig. \ref{fig:usf_D_distri}a and \ref{fig:usf_D_distri}b for the alloys without or with Re or Ru, respectively. In principle, an alloy with a larger $\gamma_{usf}$ could potentially have a higher mechanical strength because higher stress may be required for the dislocation nucleation and motion. Also, based on the Rice criterion of crack-tip deformation\cite{rice1974ductile,rice1992dislocation}, the alloys with larger $D$ parameters should have stronger potency of being intrinsic ductile. Therefore, the results of Fig. \ref{fig:usf_D_distri} provide qualitative but comprehensive evaluations of the strength-ductility balance of the 112,378 quaternary refractory alloys studied in the screening process. As shown in Fig. \ref{fig:usf_D_distri}a and \ref{fig:usf_D_distri}b, $\gamma_{usf}$ of both the alloys containing and not containing Re or Ru coarsely show negatively nonlinear correlations with the variations of the $D$ parameter. This result implies that the strength-ductility relationships in the bcc refractory alloys overall follow the classic pattern that alloys with higher mechanical strengths generally should have poorer ductility. 

Nevertheless, as shown in Fig. \ref{fig:usf_D_distri}a and b, the distributions between $\gamma_{usf}$ and the $D$ parameter are also quite dispersed. In other words, the alloys with similar USF energies can still have very different D parameters, indicating considerable deviations between their ductility performances. Therefore, there are still large degrees of freedom to tune the alloy compositions for optimal strength-ductility synergy. Apparently, the alloys corresponding to the data on the upper edge of the $\gamma_{usf}$ vs. $D$ parameter distribution should be promising to achieve better strength-ductility synergy, since they have either larger USF energies or values of the $D$ parameter compared to other alloys with similar $D$ or USF energy values, respectively. 

On the other hand, under the concept of HEAs, many of the previous experimental syntheses on the refractory multicomponent alloys are mainly focused on those with equimolar compositions. As a comparison, we also particularly marked the positions of six previously reported equimolar alloys (i.e., NbTaMoW\cite{senkov2010refractory}, VNbTaW\cite{YAO2016203}, TiZrNbMo\cite{zhang2012alloy}, TiVNbTa\cite{yang2012microstructure}, TiZrVNb\cite{senkov2013low} and TiZrHfNb\cite{wu2014refractory}) as well as three pure bcc metals (i.e. W, Mo and Nb) on the distribution plot of Fig. \ref{fig:usf_D_distri}a. As seen, most of these equimolar alloys are located away from the upper edge of the distribution in Fig. \ref{fig:usf_D_distri}a and \ref{fig:usf_D_distri}b. This result suggests that there can be other undiscovered alloy compositions, possibly deviated from equimolar, at which we may achieve comparable mechanical strengths with these currently known alloys but improved ductility. These undiscovered alloy compositions can be further rigorously located by an integration of the present surrogate models with other computational models for the predictions of their phase stability, such as the CALPHAD method\cite{liu2009first,miracle2014exploration,gao2016design} and recently developed machine-learning-based and Monte-Carlo-based models\cite{pei2020machine,zhou2019machine,huang2019machine,antillon2020efficient,natarajan2018machine}, and accurate mechanical properties\cite{maresca2020mechanistic,yin2020vanadium,george2020high,rao2017atomistic,rao2019solution}.

The effects of Re and Ru can be observed by a comparison between Fig. \ref{fig:usf_D_distri}a and Fig. \ref{fig:usf_D_distri}b. Clearly, the addition of Re or Ru leads to a more spread distribution between $\gamma_{usf}$ and the $D$ parameter, which possibly provides even larger space for optimizing the combination of strength and ductility. Additionally, as shown in Fig. \ref{fig:usf_D_distri}b, the $\gamma_{usf}$ and $D$ parameter of a part of the Re/Ru-containing alloys exhibit a different distribution pattern that bowing out from the general trend, which is resulted from the part of the $\gamma_{usf}$/$\gamma_{surf}$ distribution marked in Fig. \ref{fig:usf_surf_distri}b. Correspondingly, compared to the Re/Ru-free alloys with similar USF energies, these alloys generally have much higher values of $D$ parameter, meaning a stronger potency of being intrinsic ductile. The result suggests that the ductility of the bcc refractory multicomponent alloys can be improved without largely trading off the mechanical strength by adding proper amounts of Re or Ru. This argument is also supported by the recent experimental observations\cite{zhang2020microstructure}. On the other hand, it should also be acknowledged that the rareness of Re and Ru may make them only available for the alloy design with particular purposes.

Furthermore, it is necessary and valuable to further validate the screening results discussed above using DFT calculations. Specifically, from the alloy compositions appearing near the upper edge of the $\gamma_{usf}$ vs. $D$ parameter distributions shown in Fig. \ref{fig:usf_D_distri}, we randomly selected 25 compositions at which additional DFT calculations with the SQS method were performed to validate the predictions of the surrogate models. Additionally, to further convince the reliability of the surrogate models, we also use DFT calculations to verify a part of the predicted extreme values. More specifically, it is noticed that the surrogate model predicts a few of alloy compositions at which extremely high USF energies can be achieved, even larger than that of pure W as shown in Fig. \ref{fig:usf_D_distri}a. Therefore, from these alloy compositions with extreme $\gamma_{usf}$, we also randomly selected 6 compositions for validation. As shown in Fig. \ref{fig:usf_D_distri}a and \ref{fig:usf_D_distri}b, the locations of the selected alloy compositions on the distributions of $\gamma_{usf}$ vs. $D$ parameter are particularly marked using the solid circles with red color. 

\begin{figure}[H]
    \centering
    \includegraphics[width=1.0\linewidth]{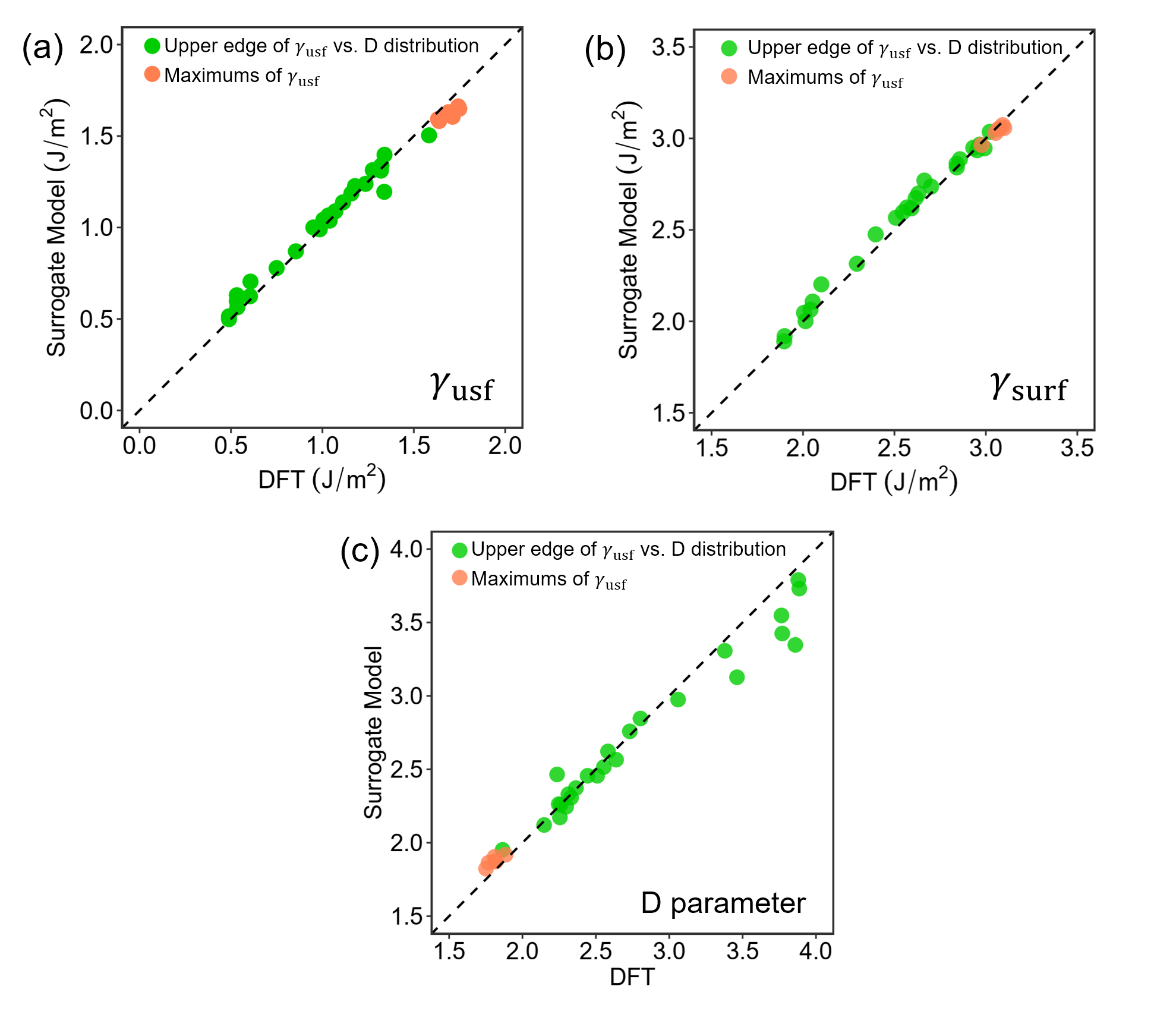}
    \caption{DFT validations of the screening results. The predictions of the surrogate models on (a) unstable stacking fault energy ($\gamma_{usf}$), (b) surface energy ($\gamma_{surf}$), and (c) $D$ parameter are plotted versus the validation results from the DFT calculations. The alloy compositions selected for validation are two sets of data. One set contains 25 samples, marked by green circles, which are randomly selected from the alloy compositions appearing near the upper edge of the $\gamma_{usf}$ vs. $D$ parameter distributions shown in Fig. \ref{fig:usf_D_distri}. The other set contains 6 samples, marked by red circles, randomly selected from the alloy compositions with extreme high $\gamma_{usf}$.} 
    \label{fig:screening_validation}
\end{figure}

The validation results are also illustrated as parity plots shown in Fig. \ref{fig:screening_validation}a, \ref{fig:screening_validation}b and \ref{fig:screening_validation}c for $\gamma_{usf}$, $\gamma_{surf}$, and the $D$ parameter, respectively. As shown in Fig. \ref{fig:screening_validation}a and \ref{fig:screening_validation}b, the predictions of the surrogate models on the USF and surface energies both agree well with the results of the DFT calculations. The corresponding RMSEs are 0.058 and 0.044 $J/m^2$, respectively. In terms of the $D$ parameter, most of the model predictions are in a good agreement with the DFT results (Fig. \ref{fig:screening_validation}c), though deviations are also observed for a few of alloys with relatively large $D$ parameters. It should be noted that the surrogate model was never trained by directly using the $D$ parameter as the fitting response. The predictions on the $D$ parameter are actually derived from the correspondingly predicted $\gamma_{usf}$ and $\gamma_{surf}$ based on Eq. \ref{Eq:D_param}. Therefore, the predicted $D$ parameters are influenced by both the prediction uncertainties of $\gamma_{usf}$ and $\gamma_{surf}$, consequently having relatively larger errors. Nevertheless, the overall good agreement between the model predictions and DFT results demonstrates the reliability of the screening results shown in Fig. \ref{fig:usf_surf_distri} and \ref{fig:usf_D_distri}. 

\section{Summary and conclusion}
In this work, we developed surrogate models based on statistical regression to effectively and efficiently predict the USF ($\gamma_{usf}$) and surface ($\gamma_{surf}$) energies of the ($\bar110$) plane in multicomponent bcc solid-solution alloys with constituent elements among Ti, Zr, Hf, V, Nb, Ta, Mo, W, Re and Ru. DFT calculations with the SQS method were performed to compute $\gamma_{usf}$ and $\gamma_{surf}$ for 106 individual alloy compositions in 14 binary, 3 ternary, and 12 quaternary systems to train and test the surrogate models. From the DFT calculations, it is also found that the variations of $\gamma_{usf}$, $\gamma_{surf}$, and their ratio in bcc refractory alloys are hardly explained in a quantitative manner solely by the d-band filling effects. Therefore, enlightened by the bond-counting model, a set of descriptors were developed to incorporate not only the filling fraction of d-band but also various features of the chemical bonds and electronic structures of pure metals and ordered intermetallic alloys for the statistical regression of $\gamma_{usf}$ and $\gamma_{surf}$. As a result, by only trained with the data of binary and ternary alloys, the developed surrogate models can accurately predict $\gamma_{usf}$ and $\gamma_{surf}$ for multicomponent alloys across a vast compositional space. The models also show the potential capability to extend predictions to cover new types of constituent elements beyond the training data. Furthermore, because of the way of descriptor constructions, the models are able to yield immediate predictions for multicomponent alloys by solely using the alloy compositions as inputs without the need of any further DFT calculations.

Moreover, using the developed surrogate models, a systematic screening of $\gamma_{usf}$, $\gamma_{surf}$ and their ratios (i.e., the $D$ parameter) were performed over 112,378 quaternary alloy compositions in the 10-element compositional space. As the potency of an alloy being mechanically strong and intrinsically ductile is generally related to its $\gamma_{usf}$ and $D=\frac{\gamma_{surf}}{\gamma_{usf}}$, respectively, the evaluation on the strength-ductility balance of bcc multicomponent refractory alloys was attempted by analyzing the screening results. The results suggest that there could be considerable spaces to tune alloy chemical compositions for further improvements of the strength-ductility synergy relative to the currently known equimolar HEAs. Besides, it is found that introducing Re or Ru can be beneficial to improve the ductility of the multicomponent alloys without largely sacrificing the mechanical strength, although the rareness of the two elements may restrict the applications in practice. Last but not least, the screening results were further confirmed by additional DFT calculations, from which some promising alloy compositions were proposed for future computational and experimental investigations towards the design of bcc refractory multicomponent alloys with enhanced strength and ductility.

\clearpage

\clearpage
\bibliographystyle{unsrt}  
\bibliography{main}
\clearpage

\begin{appendices}
\singlespacing
\sectionfont{\small}

\renewcommand*\thetable{A\arabic{table}}

\section{Bond feature parameters and elemental properties for descriptors construction}
\setcounter{table}{0} 
\footnotesize
\LTcapwidth=\textwidth

\begin{longtable}{|c|c|c|}
    \caption{A list of bond feature parameters and elemental properties used for descriptor constructions. The bond feature parameters correspond to a group of physical and electronic properties of the bcc pure metals and ordered B2 intermetallics obtained from DFT calculations.}
    \label{Table:feature}
    \endfirsthead
    \endhead
    \hline
     &  &{Parameters for descriptors construction} \\
     \hline
    \multirow{16}*{\rotatebox{90}{\tabincell{c}{Bond feature parameters}}} 
    & \multirow{12}*{\tabincell{c}{associated with \\ the bulk structure}}  & {$E_c^{bcc/b2}$: cohesive energy}\\
    \cline{3-3}
    
    &
    & $V_{eq}^{bcc/b2}$: equilibrium atomic volume \\
    \cline{3-3}
    
    &
    & \tabincell{c}{$\epsilon^1_{sp}$(bulk): first order moment of the valence $sp$-orbital \\ LDOS of the atom in perfect bulk lattice}  \\
    \cline{3-3}
    
    &
    & \tabincell{c}{$\epsilon^2_{sp}$(bulk): second order moment of the valence $sp$-orbital \\ LDOS of the atom in perfect bulk lattice}  \\
    \cline{3-3}

    & 
    & \tabincell{c}{$\epsilon^1_{d}$(bulk): first order moment of the valence $d$-orbital \\ LDOS of the atom in perfect bulk lattice}  \\
    \cline{3-3}
    
    & 
    & \tabincell{c}{$\epsilon^2_{d}$(bulk): second order moment of the valence $d$-orbital \\ LDOS of the atom in perfect bulk lattice}  \\
    \cline{3-3}
   
    & 
    & \tabincell{c}{$dip$(bulk): bimodality of the valence $d$-orbital \\ LDOS of the atom in perfect bulk lattice}  \\
    \cline{2-3}
    
    & \multirow{4}*{\tabincell{c}{associated with \\ the GSF structure}} 
    & \tabincell{c}{$\gamma_{usf}^{bcc/b2}$: the unstable stacking fault energy of \\ the single-element bcc and binary B2 structures }  \\
    \cline{3-3}
    
        &
    & \tabincell{c}{$\epsilon^1_{sp}$(USF): first order moment of the valence $sp$-orbital \\ LDOS of the atom on the fault plane}  \\
    \cline{3-3}
    
    &
    & \tabincell{c}{$\epsilon^2_{sp}$(USF): second order moment of the valence $sp$-orbital \\ LDOS of the atom on the fault plane}  \\
    \cline{3-3}

    & 
    & \tabincell{c}{$\epsilon^1_{d}$(USF): first order moment of the valence $d$-orbital \\ LDOS of the atom on the fault plane}  \\
    \cline{3-3}
    
    & 
    & \tabincell{c}{$\epsilon^2_{d}$(USF): second order moment of the valence $d$-orbital \\ LDOS of the atom on the fault plane}  \\
    \cline{3-3}
   
    & 
    & \tabincell{c}{$dip$(USF): bimodality of the valence $d$-orbital \\ LDOS of the atom on the fault plane}  \\
    \cline{2-3}
    
    & \multirow{12}*{\tabincell{c}{associated with the \\ surface structure}} 
    & \tabincell{c}{$\gamma_{surf}^{bcc/b2}$: the surface energy of \\ the single-element bcc and binary B2 structures }  \\
    \cline{3-3}
    
        &
    & \tabincell{c}{$\epsilon^1_{sp}$(Surf): first order moment of the valence $sp$-orbital \\ LDOS of the atom on the surface plane}  \\
    \cline{3-3}
    
    &
    & \tabincell{c}{$\epsilon^2_{sp}$(Surf): second order moment of the valence $sp$-orbital \\ LDOS of the atom on the surface plane}  \\
    \cline{3-3}

    & 
    & \tabincell{c}{$\epsilon^1_{d}$(Surf): first order moment of the valence $d$-orbital \\ LDOS of the atom on the surface plane}  \\
    \cline{3-3}
    
    & 
    & \tabincell{c}{$\epsilon^2_{d}$(Surf): second order moment of the valence $d$-orbital \\ LDOS of the atom on the surface plane}  \\
    \cline{3-3}
   
    & 
    & \tabincell{c}{$dip$(Surf): bimodality of the valence $d$-orbital \\ LDOS of the atom on the surface plane}  \\
    \hline 

    \multicolumn{2}{|c|}{\multirow{2}*{Elemental properties}}
    & $\chi$: electronegativity by Pauling scale \\
    \cline{3-3}
    
    \multicolumn{2}{|c|}{~}
    & $N_{val}$: number of valence electrons \\
    \hline
\end{longtable}
\end{appendices}

\end{document}